\documentclass[a4paper,11pt]{article}
\pdfoutput=1
\usepackage{jheppub} 
\usepackage[T1]{fontenc} 
\usepackage[normalem]{ulem}
\usepackage{multirow}
\usepackage{braket}
\usepackage{float}
\usepackage{setspace}
\usepackage{booktabs}
\usepackage{xcolor}
\usepackage[graphicx]{realboxes}
\usepackage{url}

\newcommand{\eg}{\textit{e.g.~}}

\title{\boldmath Global oscillation data analysis on the $3\nu$ mixing without unitarity}

\preprint{NCTS-PH/2010}
\author[a]{Zhuojun Hu,}
\author[a]{Jiajie Ling,}
\author[a]{Jian Tang,}
\author[b,a]{TseChun Wang}

\affiliation[a]{School of Physics, Sun Yat-sen University, Guangzhou 510275, China}
\affiliation[b]{National Center for Theoretical Sciences, Physics Division, Hsinchu, 30013, Taiwan}
\emailAdd{huzhj3@mail2.sysu.edu.cn}
\emailAdd{lingjj5@mail.sysu.edu.cn}
\emailAdd{tangjian5@mail.sysu.edu.cn}
\emailAdd{tsechunwang@mx.nthu.edu.tw }

\abstract{
We present results of a combined analysis in neutrino oscillations without unitarity assumption in the $3\nu$ mixing picture. Constraints on neutrino mixing matrix elements are based on recent data from the reactor, solar and long-baseline accelerator neutrino oscillation experiments. The current data are consistent with the standard $3\nu$ scheme. The precision on different matrix elements can be as good as a few percent at $3\sigma$ CL, and is mainly limited by the experimental statistical uncertainty. The $\nu_e$ related elements are the most precisely measured among all sectors with the uncertainties $<20\%$. The measured leptonic CP violation is very close to the one assuming the standard $3\nu$ mixing. The deviations on normalization and the unitarity triangle closure are confined within $\mathcal{O}(10^{-3})$, $\mathcal{O}(10^{-2})$ and $\mathcal{O}(10^{-1})$, for $\nu_e$, $\nu_{\mu}$ and $\nu_{\tau}$ sectors, respectively. We look forward to the next-generation neutrino oscillation experiments \textit{such as} DUNE, T2HK, and JUNO, especially the precise measurements on $\nu_\tau$ oscillations, to significantly improve the precision of unitarity test on the $3\nu$ mixing matrix. 
}

\begin{document} 
\maketitle
\flushbottom

\section{Introduction}\label{sec:intro}

The observation of neutrino oscillation phenomena \cite{Ahmad:2002jz,Fukuda:1998ah,Eguchi:2002dm}, which can be described by neutrino masses and the leptonic mixing, points to new physics beyond the Standard Model (SM). The leptonic mixing can be incorporated into the SM by involving the neutrino mixing matrix $U$ so that neutrino flavour eigenstates are identified by mass eigenstates with the superposition $\ket{\nu_{\alpha}} = \sum U_{\alpha i}\ket{\nu_i}$, where $U_{\alpha i}$ is an element of $U$. So far, $\nu_e$, $\nu_{\mu}$ and $\nu_{\tau}$ are three observed neutrino flavour eigenstates that participate in standard weak interactions. They are called `active neutrinos'. Without any experimental evidence of the new neutrino states in the neutrino mixing, the matrix $U$ is generally assumed to be a unitary $3\times3$ matrix $U^{3\nu}$, and is commonly parameterized by the Pontecorvo-Maki-Nakagawa-Sakata (PMNS) matrix \cite{Pontecorvo:1957cp, Pontecorvo:1967fh, Maki:1962mu} with three mixing angles $\theta_{12}$, $\theta_{13}$ and $\theta_{23}$, a Dirac phase $\delta_\text{CP}$, and two Majorana phases. In principle, two Majorana phases never affect the neutrino oscillation formalism. Therefore, we will only talk about the Dirac CP phase in the following discussion. 

In the standard $3\nu$ mixing framework, the recent data from solar, atmospheric, reactor and long-baseline accelerator neutrino oscillation experiments can be combined to constrain these neutrino mixing parameters \cite{1809173}.
Two out of four parameters in the PMNS matrix have already been measured in a relatively high precision, rather than the other two. One is the parameter $\sin^2\theta_{12}$ primarily determined by solar neutrino experiments like SNO and the long-baseline reactor experiment like KamLAND. The other is $\sin^2\theta_{13}$, reaching a precision of $\sim3\%$, predominantly derived from the reactor neutrino experiments such as Daya Bay, Double Chooz and RENO, with baselines at $\sim1$ km. Apart from the current achievement, there are still puzzles in neutrino physics to be resolved: $\theta_{23}$-octant problem ($\theta_{23}>,~=$ or $<45^\circ$), the neutrino mass ordering problem (if $\Delta m_{31}^2>$ or $<0$), whether there is any CP violation in the leptonic sector and the size of $\delta_\text{CP}$ if so. According to the current result, it seems that, by the combining atmospheric and accelerator neutrino oscillation data, we will have $\sin^2\theta_{23}$ in the higher octant and the CP violation scenario $\delta_\text{CP}\ne 0,\pi$. Recent updates from running accelerator neutrino experiments, NOvA and T2K \cite{Acero:2019ksn, Abe:2019vii} have significantly improved the precision of these two parameters. Given the standard $3\nu$ mixing, the current global fit results prefer the solution with $\sin^2\theta_{12}=0.307^{+0.013}_{-0.012}$, $\sin^2\theta_{23}=0.542^{+0.019}_{-0.022}$ (normal ordering, upper octant) and $\sin^2\theta_{13}=0.0218\pm0.0007$. In addition, the remaining parameter $\delta_\text{CP}$ shows an inclination to be non-zero, though the uncertainty $\Delta \delta_\text{CP}$ still needs to be reduced \cite{Tanabashi:2018oca}.

Entering a new era of precise measurements in neutrino oscillation physics, whether the $3\nu$ mixing framework conserves unitarity deserves further scrutiny. Moreover, our understanding of neutrino mixing does not necessitate the foundation of unitary $3\nu$ mixing assumption. There are theoretical extensions in the neutrino sector in the SM, which allow the non-unitarity of $3\nu$ mixing, \textit{e.g.} the sterile-active neutrino mixing \cite{Langacker:1988up, Czakon:2001em, Bekman:2002zk, Loinaz:2003gc, Malinsky:2009df, Antusch:2009gn, Forero:2011pc, Berryman:2014yoa,Dev:2009aw}. The motivations for this new mixing are (1) to explain experimental anomalies~\cite{Mention:2011rk,Aguilar:2001ty,Aguilar-Arevalo:2018gpe}, (2) to explain the origin of neutrino masses~\textit{e.g.,} \textbf{seesaw I} \cite{Minkowski:1977sc, yanagida1979proceedings, freedman1982viii, Mohapatra:1979ia} and \textbf{seesaw III} \cite{Foot:1988aq, Ma:1998zg}, and (3) to explain warm dark matter~\cite{Asaka:2005pn, Baumholzer:2018sfb, Boyarsky:2009ix}. In addition, unknown couplings involving neutrinos may result in effective the $3\nu$ non-unitarity\cite{Antusch:2008tz, Meloni:2009cg, Ohlsson:2012kf}. As the precision gets better and better, one might be always questioning how much we know about the neutrino mixing if taking away the unitarity assumptions, and whether the unitarity assumption is valid. In the quark sector, similar issues have been discussed for decades, \textit{such as}~Refs.~\cite{Charles:2004jd,Hocker:2001xe,Bona:2006ah,CKMfitter,Akama:1991tk,Chen:2007yn}. Some of the neutrino phenomenological studies answer these questions~\cite{Antusch:2006vwa, Escrihuela:2015wra,Fong:2016yyh,Parke:2015goa,Ellis:2020ehi,Ellis:2020hus}. Along with Ref.~\cite{Parke:2015goa}, testing the unitarity hypothesis in a general manner, authors of current works~\cite{Ellis:2020ehi,Ellis:2020hus} have updated the results with the current data, and their predicted results emphasis the prospect of future experiments: DUNE, Hyper, and JUNO, as well as the $\tau$ neutrino measurements.

In this paper, we present an analysis beyond the standard $3\nu$ mixing scheme, using current neutrino oscillation data. In presence of sterile neutrinos, the neutrino mixing matrix is enlarged to a unitary $(3+N)\times(3+N)$ mixing matrix, where there are $N$ types of sterile neutrinos. This results in the non-unitarity of the $3\nu$ mixing matrix. The non-unitary $3\nu$ mixing assumption is compared against data from accelerator, reactor and solar neutrino measurements. Our understanding of the $3\nu$ mixing matrix is so far mostly limited to the $\nu_e$ and $\nu_{\mu}$ sectors, by reactor and solar neutrino experiments and accelerator neutrino experiments, respectively. In our analysis, we will try to answer how the neutrino mixing looks like beyond the standard $3\nu$ mixing scheme and how much the current data prefer the $3\nu$ unitarity assumption. 

This paper is organized as follows. In section \ref{sec:model} we introduce the non-unitary $3\nu$ mixing framework. Section \ref{sec:setting} describes the oscillation data used in this analysis, as well as how we analyse the data. Section \ref{sec:result} presents the resultant non-unitary $3\nu$ mixing parameters, their correlations and CP-violation, as well as the results of unitarity test. We present allowed parameter regions of the $e-\mu$ triangle in Fig.~\ref{fig:Triangle_1} without assuming unitarity. Unitarity conditions are tested and shown in Fig.~\ref{fig:unitarity_condition_normalization}. Finally, we summarize and conclude this work in section \ref{sec:conclusion}.

\section{Non-unitarity and neutrino oscillation}\label{sec:model}

The neutrino oscillation is a quantum coherent phenomenon described by the neutrino mixing and the mass-squared differences. The presence of sterile neutrinos in neutrino oscillations requires extra neutrino states and the extension of the mixing matrix from $3\times3$ to the larger. Assuming the system with three active neutrinos and $N$ sterile neutrinos, the whole $3+N$ mixing matrix as shown in Eq.~(\ref{eq:3+N_mixing}) is a complete and unitary mixing matrix. As a result, the left-up $3\times3$ subset that couples $e, \mu, \tau$ and $\nu_1, \nu_2, \nu_3$ is non-unitary. The effect of non-unitarity is detectable, for example, deficits in disappearance measurements as a part of active neutrinos change into invisible sterile neutrinos.

\begin{equation}\label{eq:3+N_mixing}
  \begin{pmatrix}
  \nu_e \\
  \nu_{\mu} \\
  \nu_{\tau} \\
  \nu_{s} \\
  \vdots
  \end{pmatrix}
  =
  \begin{pmatrix}
    U_{e1} & U_{e2} & U_{e3} & U_{e4} & \cdots\\
    U_{\mu 1} & U_{\mu 2} & U_{\mu 3} & U_{\mu 4} & \cdots\\
    U_{\tau 1} & U_{\tau 2} & U_{\tau 3} & U_{\tau 4} & \cdots \\
    U_{s1} & U_{s2} & U_{s3} & U_{s4} & \cdots \\
    \cdots & \cdots & \cdots & \cdots & \ddots 
  \end{pmatrix}
  \begin{pmatrix}
  \nu_1 \\
  \nu_2 \\
  \nu_3 \\
  \nu_4 \\
  \vdots
  \end{pmatrix}.
\end{equation}

In this section, we will introduce the standard $3\nu$ scheme and the basic conditions for its unitarity assumption. Then a general $3\nu$ mixing matrix relaxing the unitarity conditions will be presented. Additional conditions in the non-unitarity assumption will be also discussed. After presenting the way to discuss the CP violation, we will come up to discuss neutrino oscillations under the assumption of non-unitarity.

\subsection{The $3\times 3$ unitary mixing matrix}

In the standard three-neutrino-mixing scheme, the mixing matrix $U^{3\nu}$ between the flavour and mass eigenstates is defined as
\begin{equation}\label{eq:3nu_mixing}
  \begin{pmatrix}
  \nu_e \\
  \nu_{\mu} \\
  \nu_{\tau}
  \end{pmatrix}
  =
  \begin{pmatrix}
    U_{e1} & U_{e2} & U_{e3} \\
    U_{\mu 1} & U_{\mu 2} & U_{\mu 3} \\
    U_{\tau 1} & U_{\tau 2} & U_{\tau 3} 
  \end{pmatrix}
  \begin{pmatrix}
  \nu_1 \\
  \nu_2 \\
  \nu_3 \\
  \end{pmatrix}.
\end{equation} 
Neglecting the two Majorana CP phases, which have no effect on neutrino oscillations, $U^\text{PMNS}$ is given by three-dimensional rotation matrix described with three mixing angles and one Dirac CP phase,
\begin{equation}
U^\text{PMNS}=  \begin{pmatrix}
    1 & 0 & 0 \\
    0 & \cos \theta_{23} & \sin\theta_{23} \\
    0 & -\sin\theta_{23} & \cos\theta_{23} 
  \end{pmatrix}\cdot
  \begin{pmatrix}
    \sin\theta_{13} & 0 & \sin\theta_{13}\mathrm{e}^{-i\delta_\text{CP}} \\
    0 & 1 & 0 \\
    -\sin\theta_{13}\mathrm{e}^{i\delta_\text{CP}} & 0 & \cos\theta_{13} 
  \end{pmatrix}\cdot
  \begin{pmatrix}
    \cos\theta_{12} & \sin\theta_{12} & 0 \\
    -\sin\theta_{12} & \cos \theta_{12} & 0 \\
    0 & 0 & 1 
  \end{pmatrix},
\end{equation}
where the $\theta_{ij}$ is the mixing angle for each rotation, and $\delta_\text{CP}$ is the Dirac CP phase. This parametrization satisfies the unitarity conditions,

  \begin{align}
    |U^{3\nu}_{\alpha 1}|^2 + |U^{3\nu}_{\alpha 2}|^2 + |U^{3\nu}_{\alpha 3}|^2 &= 1,~~\alpha=e,\mu,\tau, \label{eq:unitarity_1}
    \\
    |U^{3\nu}_{e i}|^2 + |U^{3\nu}_{\mu i}|^2 + |U^{3\nu}_{\tau i}|^2 &= 1,~~i=1,2,3, \label{eq:unitarity_2}
    \\
    U^{3\nu}_{\alpha 1}U^{3\nu,*}_{\beta 1} + U^{3\nu}_{\alpha 2}U^{3\nu,*}_{\beta 2} + U^{3\nu}_{\alpha 3}U^{3\nu,*}_{\beta 3} &= 0,~~\alpha, \beta = e,\mu,\tau,~~\alpha\neq\beta, \label{eq:unitarity_3}
    \\
    U^{3\nu}_{e i}U^{3\nu,*}_{e j} + U^{3\nu}_{\mu i}U^{3\nu,*}_{\mu j} + U^{3\nu}_{\tau i}U^{3\nu,*}_{\tau j} &= 0,~~i,j  =1,2,3,~~i\neq j. \label{eq:unitarity_4}
  \end{align}

We note that the first two conditions Eq.~(\ref{eq:unitarity_1}) and (\ref{eq:unitarity_2}), are the normalization conditions, while the other two Eq.~(\ref{eq:unitarity_3}) and (\ref{eq:unitarity_4}) are for the unitarity triangle closure. These conditions break in the $(3+N)$-neutrino scenario. For example, in the $3+1$ case, $|U_{\alpha 1}|^2 + |U_{\alpha 2}|^2 + |U_{\alpha 3}|^2 = 1-|U_{\alpha 4}|^2 <1$. As a result, by testing the validity of these conditions, we can exam the standard $3\nu$ scheme.

\subsection{The $3\times 3$ non-unitary mixing matrix}

A general $3\times3$ mixing matrix consists of nine complex matrix elements. Each of them has real and imaginary parts. In total, there are 18 parameters in this matrix. Among them, five phases can be removed by redefining the phase in eigenstates, leaving nine absolute values and four phases, and the physics is unchanged. One possible parametrization is given as 
\begin{equation}\label{eq:3nu_non_uni_par}
U^{NU} = 
\begin{pmatrix}
|U_{e1}|e^{i\phi_{e 1}} & |U_{e2}|e^{i\phi_{e 2}} & |U_{e3}| \\
|U_{\mu 1}|e^{i\phi_{\mu 1}} & |U_{\mu 2}|e^{i\phi_{\mu 2}} & |U_{\mu 3}| \\
|U_{\tau 1}| & |U_{\tau 2}| & |U_{\tau 3}|
\end{pmatrix}.
\end{equation}
The assignment of the four phases (or image parts) can be arbitrary, while this makes no difference in neutrino oscillations. We use such parametrization Eq.~(\ref{eq:3nu_non_uni_par}) in the following analysis. 

Once the neutrino mixing goes beyond the standard $3\nu$ scheme, equalities in Eqs.~(\ref{eq:unitarity_1})-(\ref{eq:unitarity_4}) are not preserved. 
Given $U^{NU}$ is a $3\times3$ sub-matrix of a $(3+N)\times(3+N)$ unitary matrix, the sum of squares of each row or column in $U^{NU}$ shall never exceeds 1. Thus the $3\times3$ matrix $U^{NU}$ satisfies:
\begin{equation}
  \begin{aligned}
    |U^{NU}_{\alpha 1}|^2 + |U^{NU}_{\alpha 2}|^2 + |U^{NU}_{\alpha 3}|^2 &\le 1,~~\alpha=e,\mu,\tau;
    \\
    |U^{NU}_{e i}|^2 + |U^{NU}_{\mu i}|^2 + |U^{NU}_{\tau i}|^2 &\le1,~~i=1,2,3.
  \end{aligned}
\label{eq:geometric_leq1}
\end{equation}

Considering any non-unitarity in $U^{NU}$ induced by the active-sterile mixing, additional constraints can be applied, to further bound the parameter space. As discussed in Ref.~\cite{Parke:2015goa,Fong:2016yyh}, the active-sterile mixing matrix elements in the $(3+N)\times(3+N)$ matrix satisfy Cauchy-Schwarz inequalities
\begin{align}
  \left|\sum^{N+3}_{i=4}U^{NU}_{\alpha i}U_{\beta i}^{NU,*}\right|^{2} \leq & \left(\sum^{N+3}_{i=4}|U^{NU}_{\alpha i}|^{2}\right) \left(\sum^{N+3}_{i=4}|U^{NU}_{\beta i}|^{2}\right),
   &\mathrm{for}~\alpha, \beta = e, \mu, \tau, \alpha \neq \beta, \label{eqn:Cauchy-Schwarz-Sterile_1}
  \\
  \left|\sum^{s^{N}}_{\alpha=s^{1}}U^{NU}_{\alpha i}U_{\alpha j}^{NU,*}\right|^{2} \leq & \left(\sum^{s^{N}}_{\alpha=s^{1}}|U^{NU}_{\alpha i}|^{2}\right) \left(\sum^{s^{N}}_{\alpha=s^{1}}|U^{NU}_{\alpha j}|^{2}\right),
   &\mathrm{for}~i, j = 1, 2, 3, i \neq j, \label{eqn:Cauchy-Schwarz-Sterile_2}
\end{align}
where $s^{k}$ denotes the $k$th sterile state. By noticing the unitarity of the $(3+N)\times(3+N)$ matrix, Eqs.(\ref{eqn:Cauchy-Schwarz-Sterile_1}) and (\ref{eqn:Cauchy-Schwarz-Sterile_2}) can be rewritten as
\begin{align}
  \left|\sum^{3}_{i=1}U^{NU}_{\alpha i}U_{\beta i}^{NU,*}\right|^{2} \leq & \left(1-\sum^{3}_{i=1}|U^{NU}_{\alpha i}|^{2}\right) \left(1-\sum^{3}_{i=1}|U^{NU}_{\beta i}|^{2}\right),
  &\mathrm{for}~\alpha, \beta = e, \mu, \tau, \alpha \neq \beta, \label{eqn:Cauchy-Schwarz-Active_1}
  \\
  \left|\sum^{\tau}_{\alpha=e}U^{NU}_{\alpha i}U_{\alpha j}^{NU,*}\right|^{2} \leq & \left(1-\sum^{\tau}_{\alpha=e}|U^{NU}_{\alpha i}|^{2}\right) \left(1-\sum^{\tau}_{\alpha=e}|U^{NU}_{\alpha j}|^{2}\right),
   &\mathrm{for}~i, j = 1, 2, 3, i \neq j. \label{eqn:Cauchy-Schwarz-Active_2}
\end{align}

We note here that the inequalities Eqs.~(\ref{eqn:Cauchy-Schwarz-Active_1}) and (\ref{eqn:Cauchy-Schwarz-Active_2}) need to be satisfied as requiring at least three active neutrinos join in the neutrino mixing. As a result, we will impose these inequality conditions in our later analysis. We need to keep in mind that the constraint on each matrix element of $U^{NU}$ is expected to be worse than those for $U^{3\nu}$. This is not only because the degrees of freedom are extended for $U^{NU}$, but also the restrictions on these 9 elements are weaker (compare Eqs.~(\ref{eqn:Cauchy-Schwarz-Active_1},\ref{eqn:Cauchy-Schwarz-Active_2}) to Eqs.~(\ref{eq:unitarity_1}-\ref{eq:unitarity_4})). In addition, it is also interesting to ask if more degeneracy solutions will appear in $U^{NU}$, or if the existing degeneracy solutions in $U^{3\nu}$ is harder to be separated from the current best fit. These will be included in our later analysis.

\subsection{CP violation}

The CP violation can happen in the active-neutrino sector, and is defined as
\begin{equation}\label{eq:CPV}
U_{3\times3}\ne U^*_{3\times3},
\end{equation}
where $U_{3\times3}$ can be $U^{3\nu}$ or $U^{NU}$. The \textbf{size} of CP violation for neutrino oscillations is described by the Dirac CP phase $\delta_\text{CP}$ in the standard $3\nu$ scheme. As PMNS parametrization is not adoptable in $U^{NU}$, we should use a more general way to scale this \textbf{size} in the case of $3\nu$ non-unitarity. 

Under the $3\nu$ unitarity assumption, a rephasing-invariant quantity, so-called Jarlskog factor, describes the CP violation in the neutrino sector, and is the imaginary part of the four-element combination,
\begin{equation}\label{eq:jarlskog inv}
  J_{\alpha\beta i j} \equiv \Im[U_{\alpha i}U_{\beta j}U^{*}_{\alpha j}U^{*}_{\beta i}]=J\sum_\gamma\epsilon_{\alpha\beta\gamma}\sum_{k}\epsilon{ijk}.
\end{equation}
This is obvious that once the Jarlskog factor is non-zero, the CP is violated as Eq.~\ref{eq:CPV}.
The Jarlskog factor is an invariant in the standard $3\nu$ scheme, and the factor $J$ can be expressed,
\begin{equation}
J=\sin\theta_{13}\cos^2\theta_{13}\sin\theta_{12}\cos\theta_{12}\sin\theta_{23}\cos\theta_{23}\sin\delta.
\end{equation}
As our analysis is beyond the standard $3\nu$ scheme, Jarlskog factors do not need to be the same for different combinations. In other words, the second equality in Eq.~(\ref{eq:jarlskog inv}) does not hold. In the $\nu_{e}-\nu_{\mu}$ sector we have three Jarlskog factors: $J_{e\mu12} \equiv \Im[U_{e1}U_{\mu2}U^{*}_{e2}U^{*}_{\mu1}]$, $J_{e\mu23} \equiv \Im[U_{e2}U_{\mu3}U^{*}_{e3}U^{*}_{\mu2}]$ and $J_{e\mu13} \equiv \Im[U_{e1}U_{\mu3}U^{*}_{e3}U^{*}_{\mu1}]$, which will be discussed in the following analysis. 

\subsection{Oscillation probabilities}

With the non-unitary $3\nu$ mixing matrix $U^{NU}$, the probability of a neutrino of flavour $\alpha$ to be detected as a neutrino of flavour $\beta$ is given by
\begin{equation}
\begin{aligned}
P^{NU}_{\nu_{\alpha}\rightarrow\nu_{\beta}}&=|\sum_{i=1}U^{*}_{\beta i}U_{\alpha i}|^{2}-4\sum_{i<j}\Re{\left(U_{\alpha i}U_{\beta j}U^{*}_{\alpha j}U^{*}_{\beta i}\right)\sin^2\left(\frac{\Delta m^2_{ji}L}{4E_{\nu}}\right)} \\
&\pm 2\sum_{i<j}\Im \left( U_{\alpha i}U_{\beta j}U^{*}_{\alpha j}U^{*}_{\beta i} \right) \sin \left( \frac{\Delta m_{ji}^{2}L}{2E_{\nu}} \right),
\end{aligned}
\label{eqn:oscprob}
\end{equation}
where $\Delta m^2_{ji}$ is the mass-squared difference between $\nu_{j}$ and $\nu_{i}$. $L$ is the baseline distance and $E_{\nu}$ is the neutrino energy. The second term is a flavour-changing term, which is a function of $\Delta m^2_{ji}L/4E_{\nu}$, and is independent of CP violation. The last term, which is a combination of the imaginary part of the quartic product $( U_{\alpha i}U_{\beta j}U^{*}_{\alpha j}U^{*}_{\beta i} )$ and a periodic oscillation, is the CP violating term. It vanishes when CP is conserved. For antineutrinos, the oscillation probability is the same as Eq.~(\ref{eqn:oscprob}), but the matrix elements need to be replaced by their complex conjugate partners ($U_{\alpha i}\rightarrow U^*_{\alpha i}$). Eq.~(\ref{eqn:oscprob}) also shows that for the purpose of measuring a certain matrix element we need to choose a proper $L/E_\nu$ setup, and even the flavour of neutrino is also an important factor in the measurement. Therefore, we will include data from different experiments in the analysis, trying to constrain all elements in $U^{NU}$, including real and imaginary parts. These experiments will be introduced in the next section.

\section{Simulation details and analysis setup}\label{sec:setting}

To analyse each of the matrix elements in $U^{NU}$, we include data from a variety of neutrino oscillation experiments, \textit{such as} the Daya Bay \cite{Adey:2018zwh}, Double Chooz \cite{DoubleChooz:2019qbj}, KamLAND \cite{Gando:2010aa}, NOvA \cite{Acero:2019ksn}, OPERA \cite{Agafonova:2015jxn}, RENO \cite{Bak:2018ydk}, SNO \cite{Ahmad:2002jz, Aharmim:2005gt, Aharmim:2008kc}, and T2K experiments \cite{Abe:2019vii}. Extra results from sterile neutrino searches are also taken into account. In the following of this section, we will introduce how the neutrino mixing parameters are determined by different types of experiments, and present simulation details for the reactor, accelerator, and solar experiments. We will finally show how we combine data from all of the experiments in our statistical analysis, and how we include the extra information from sterile-neutrino-searching experiments.

\subsection{Roles of each experiments}

\begin{table}[h!]
\label{tab:Experiments_Probability}
  \caption{All relevant categories, experiments, and the corresponding measurements in this work. The abbreviations MBL and LBL denote medium and long baseline, respectively.}
  \begin{center}
  \begin{tabular}{c|c|c}
    \toprule
    Types  &  Exps  &  Measurements  \\
    \hline\hline
    \multirow{2}*{MBL Reactor}  &  RENO, Daya Bay  &  \multirow{2}*{$4|U_{e 3}|^2(|U_{e 1}|^2+|U_{e 2}|^{2})$}  \\
    & Double Chooz & \\
    \hline
    LBL Reactor & KamLAND  &  $4|U_{e 1}|^2|U_{e 2}|^2$  \\
    \hline
    Solar  &  SNO  &   $|U_{e2}|^2$  \\
    \hline
    LBL Accelerator  &  \multirow{2}*{NOvA, T2K}  &  \multirow{2}*{$4|U_{\mu 3}|^2(|U_{\mu 1}|^2+|U_{\mu 2}|^{2})$} \\
    ($\nu_{\mu}\rightarrow\nu_{\mu}$)  &  ~  &  ~ \\
    \hline
    LBL Accelerator  &  \multirow{2}*{NOvA, T2K}  &  \multirow{2}*{$4\Re [U_{e3}U_{\mu3}^{*}(U_{e1}U_{\mu1}^{*}+U_{e2}U_{\mu2}^{*})]$} \\
    ($\nu_{\mu}\rightarrow\nu_{e}$)  &  ~  &  ~ \\
    \hline
    LBL Accelerator  &  \multirow{2}*{OPERA}  &  \multirow{2}*{$4\Re [U_{\tau3}U_{\mu3}^{*}(U_{\tau1}U_{\mu1}^{*}+U_{\tau2}U_{\mu2}^{*})]$} \\
    ($\nu_{\mu}\rightarrow\nu_{\tau}$)  &  ~  &  ~ \\
    \bottomrule
  \end{tabular}
  \end{center}
\end{table}

The purpose of combining different experiments is to try to measure all matrix elements of $U^{NU}$ independently as much as possible. These experiments are sensitive to different $U^{NU}$ elements, because of measuring different oscillation channels, focusing on neutrino energy ranges or using different oscillation baselines. We classify them according to their measurements. Moreover, to better understand how many matrix elements are measured independently, we introduce these experiments in this classification method. We summarise all experiments in Table~\ref{tab:Experiments_Probability}. More details are introduced in the following.

\textbf{Medium-baseline (MBL) reactor neutrino experiments}: Medium-baseline reactor neutrino experiments play crucial roles in examining the unitarity condition in the $\nu_e$ sector. We include latest data from Daya Bay, Double Chooz, and RENO. With the baseline of $\sim1$ km and $0.8-12$ MeV neutrino energy, these experiments are sensitive to the $\bar{\nu}_e \rightarrow \bar{\nu}_e$ oscillation driven by mass squared splittings $|\Delta m^2_{31}|$ and $|\Delta m^2_{32}|$. Thus these data can be used to constrain $|U_{e3}|$, and $|U_{e1}|^2+|U_{e2}|^2$.

\textbf{Long-baseline (LBL) reactor neutrino experiments}: KamLAND is the only long-baseline reactor neutrino experiment by far. KamLAND studies the reactor anti-neutrino oscillation, utilizing more than 50 nuclear power reactors. The flux-weighted average baseline to the reactors is $\sim180$~km. The leading term of the KamLAND measurement is $4|U_{e1}|^2|U_{e2}|^2$. 

\textbf{Solar neutrino experiments}: Solar neutrino experiments study $\nu_e$ flavour conversions from the sun. In this analysis, we consider the observations of solar $^8B$ neutrinos. Generated at the center of the sun, solar $^8B$ neutrinos travel through the dense matter to the sun surface. The high matter density in the solar core enables solar $^8B$ neutrinos with energy of $\sim$6~MeV to undergo MSN resonant transitions. 
Also, in such environment with high matter density, the superposition of $\nu_e$ in the effective mass states is directly $\nu_2$. Then, some of these neutrinos travel through the vacuum to the earth. The distance between the sun and the earth is so long that the oscillation probability average out at the surface of the earth. Thus, the fraction of solar $^8B$ neutrinos that remain $\nu_e$ at the earth is the effectively the same as the fraction of solar $^8B$ neutrinos being $\nu_e$ at the surface of the sun. Measurements of solar $^8B$ neutrino charged-current flux verify this fraction, directly probe $|U_{e2}|^2$, and therefore disentangle the degeneracy between $|U_{e1}|$ and $|U_{e2}|$.

\textbf{Long-baseline accelerator neutrino experiments (NOvA and T2K)}: The non-unitarity variables $1-|U_{\mu1}|^2-|U_{\mu2}|^2-|U_{\mu3}|^2$ and $U_{e1}U_{\mu1}^{*}+U_{e2}U_{\mu2}^{*}+U_{e3}U_{\mu3}^{*}$ are mainly constrained by accelerator experiments measuring $\nu_{\mu}(\bar{\nu}_{\mu})$ disappearance channels and $\nu_{e}(\bar{\nu}_e)$ appearance channels. We consider here the NOvA and T2K experiments, which are designed to study accelerator neutrino oscillations driven by $|\Delta m^2_{31}|$ and $|\Delta m^2_{32}|$. Neglecting sub-leading terms and matter effects, their disappearance measurements and appearance measurements constrain $4|U_{\mu3}|^2(|U_{\mu1}|^2+|U_{\mu2}|^2)$ and $4\Re[U_{e3}U_{\mu3}^{*}(U_{e1}U_{\mu1}^{*}+U_{e2}U_{\mu2}^{*})]$, respectively. An ambiguity presents in measuring $|U_{\mu3}|$ and $|U_{\mu1}|^2+|U_{\mu2}|^2$, as known as "octant degeneracy" in the context of $3\nu$ unitarity \cite{Fogli:1996pv}. 

\textbf{Long-baseline accelerator neutrino experiments (OPERA)}: To get the independent measurement in the $\nu_\tau$ sector, we further include the observation of 5 $\nu_{\tau}$ events from a $\nu_{\mu}$ beam at OPERA. With a baseline of $730$~km and a beam energy peaked at $\sim20$~GeV, the experiment can only measure the tail of the first cycle of the oscillation driven by $|\Delta m^2_{31}|$ and $|\Delta m^2_{32}|$.
This measurement can be used to constrain the combination $4\Re[U_{\mu3}U_{\tau3}^{*}(U_{\mu1}U_{\tau1}^{*}+U_{\mu2}U_{\tau2}^{*})]$. 

\subsection{Medium-baseline reactor neutrino experiments}

The Daya Bay, Double Chooz, and RENO experiments adopted relative near-far measurements, allowing inter-detector ratio fits, so that uncertainties and potential biases from neutrino sources are minimized. We take the far-to-near ratios $R^{obs}_{i}$ from Double Chooz\cite{DoubleChooz:2019qbj} and RENO\cite{Bak:2018ydk}, and the measured $\bar{\nu}_e$ survival probability versus $L_{eff}/\left<E_{\nu}\right>$ from Daya Bay\cite{Ochoa:2018late}, along with the corresponding error $\sigma^{obs}_{i}$.\footnote{Data taken from (1) Daya Bay: Right figure in Page 16 in Ref.~\cite{Ochoa:2018late}; (2) Double Chooz: Fig.~4 left panel in Ref.~\cite{DoubleChooz:2019qbj}; (3) RENO: Fig.~2 in Ref.~\cite{Bak:2018ydk}.} $L_{eff}$ is the effective propagation distance at Daya Bay, and $\left<E_{\nu}\right>$ is the average true $\bar{\nu}_e$ energy. Background has been removed from these data points. 

The data points from Double Chooz and RENO are ratios of far spectra to expected no-oscillation spectra, where the expected no-oscillation spectra are obtained by weighting the observed spectra in the near detectors with no-oscillation assumptions. Noting correlated terms cancel out for a relative measurement, our prediction is given by:
\begin{equation}
    R^{pred}_{i} = \frac{\sum\limits^{\mathrm{reactor}}_{j=1}\sum\limits^{\mathrm{detector}}_{k=1}P^{NU}_{\bar{\nu}_e\rightarrow\bar{\nu}_e}\left(L_{jk}/E_{i}\right)\Phi_{j}/L^2_{jk}}{\sum\limits^{\mathrm{reactor}}_{j=1}\sum\limits^{\mathrm{detector}}_{k=1}P^{NU}_{\bar{\nu}_e\rightarrow\bar{\nu}_e}(0)\Phi_{j}/L^2_{jk}}.
    \label{eq:RENO_pred}
\end{equation}
$P^{NU}_{\bar{\nu}_e\rightarrow\bar{\nu}_e}\left(L_{jk}/E_{i}\right)$ is $\bar{\nu}_e$ survival probability Eq.~(\ref{eqn:oscprob}), given distance $L_{jk}$ and neutrino energy $E_{i}$. $\Phi_{j}$ is the number of neutrino generated in the $j$th reactor, and $L_{jk}$ denotes the distance from the $j$th reactor to the $k$th detector\footnote{Reactor information and baseline information for (1) Double Chooz: in Ref.~\cite{DoubleChooz:2019qbj}; (2) RENO: in Ref.~\cite{Ahn:2010vy}}. 

The Daya Bay data points are measured survival probabilities, and can be expressed as
\begin{equation}
  R^{obs}_i = P^{obs}_{\bar{\nu}_e\rightarrow\bar{\nu}_e}(L_{eff}/\left<E_{\nu}\right>) = \frac{N^{obs}_{i}}{N^{no-osc}_{i}},
\end{equation}
where $N^{obs}_{i}$ is the observed number of event in bin $i$. $N^{no-osc}_{i}$ is the expected number of event with no oscillation, derived from near-site measurements. We redefine $\omega_{i} \equiv (L/E)_{i}$ and use the following equation for Daya Bay prediction

\begin{equation} 
    R^{pred}_{i} = \frac{P^{NU}_{\bar{\nu}_e\rightarrow\bar{\nu}_e}\left(\omega_{i}\right)}{P^{NU}_{\bar{\nu}_e\rightarrow\bar{\nu}_e}\left(0\right)}.
    \label{eq:DYB_pred}
\end{equation}
A $\chi^2$ quantity is constructed for a medium-baseline reactor neutrino measurement, comparing our predictions with the observed ratios,
\begin{equation}
  \chi^{2}_{MR}=\sum_{i}\frac{\left(R^{pred}_{i}-R^{obs}_{i}\right)^{2}}{\left(\sigma^{obs}_{i}\right)^{2}},
  \label{eqn:reactor-mid chi2}
\end{equation}
where the subscript $_{MR}$ denotes different medium-baseline reactor experiments. All the three measurements share the same formula. Eventually $\chi^2$s for each measurements will be summed up,
\begin{equation}
  \chi^{2}_{ALL~MR}=\chi^{2}_{Daya~Bay}+\chi^{2}_{Double~Chooz}+\chi^{2}_{RENO}.
  \label{eqn:reactor-mid total}
\end{equation}

\subsection{Long-baseline reactor neutrino experiment}

KamLAND experiment uses $1$ kton of liquid scintillator to monitor $\bar{\nu}_e$ flux from more than $50$ nuclear power reactors at long baselines. We take the information of $23$ major reactors from Ref.~\cite{EnomotoSanshiro2005neutrino}, including their thermal power and distances to KamLAND detector. Our analysis is using the ratio of data to no-oscillation expectation (Fig.~5 in Ref.~\cite{Gando:2010aa}) from KamLAND. As there is only one detector, correlated terms do not cancel out. To analyse KamLAND data, we use a function similar to Eq.~\ref{eq:RENO_pred} to predict the ratio, with an additional normalization factor $\theta_{\mathrm{KLD}}$ which deals with the correlated uncertainties,
\begin{equation}
     R^{pred}_{i} = (1+\theta_{\mathrm{KLD}})\frac{\sum\limits^{\mathrm{reactor}}_{j=1}\sum\limits^{\mathrm{detector}}_{k=1}P^{NU}_{\bar{\nu}_e\rightarrow\bar{\nu}_e}\left(L_{jk}/E_{i}\right)\Phi_{j}/L^2_{jk}}{\sum\limits^{\mathrm{reactor}}_{j=1}\sum\limits^{\mathrm{detector}}_{k=1}P^{NU}_{\bar{\nu}_e\rightarrow\bar{\nu}_e}(0)\Phi_{j}/L^2_{jk}}.
    \label{eq:KamLAND_pred}
\end{equation}
Then we construct the following $\chi^2_{LR}$ function, with the uncertainty of $\theta_{\mathrm{KLD}}$ being $\sigma_{\mathrm{KLD}} = 5\%$. The notation $\sigma^{obs}_{i}$ is the published uncertainty of the corresponding observed data. 
\begin{equation}
  \chi^{2}_{LR}=\sum_{i}\frac{\left(R^{pred}_{i}-R^{obs}_{i}\right)^{2}}{\left(\sigma^{obs}_{i}\right)^{2}} + \frac{\theta_{\mathrm{KLD}}^2}{\sigma_{\mathrm{KLD}}^2}.
  \label{eqn:reactor-long chi2}
\end{equation}

\subsection{Solar neutrino experiment}

For solar neutrinos, we include the experimental results from the SNO experiment. The SNO $^8B$ solar neutrino flux measured with charged-current (CC) interactions, $\Phi^{obs}_{CC}$, are analysed\footnote{The SNO $^8B$ solar neutrino flux measured with neutral-current interaction is used to bound the non-unitarity $1-|U_{\tau1}|^2-|U_{\tau2}|^2-|U_{\tau3}|^2$ (Table.~\ref{tab:Sterile Data Summary}). Therefore we do not use the ratio of CC to neutral current (NC) fluxes, to avoid double counting.}. We take $\Phi^{obs}_{CC}$ from all three phases of the SNO experiments in Ref.~\cite{Ahmad:2002jz, Aharmim:2005gt, Aharmim:2008kc}, alongside the published errors $\sigma_{CC}$. Assuming adiabatic evolution, the predicted $^8B$ solar neutrino CC flux $\Phi^{pred}_{CC}$ is expressed as

\begin{equation}
  \Phi^{pred}_{CC} = \Phi_{^{8}B}(|\widetilde{U_{e1}}|^2|U_{e1}|^2+|\widetilde{U_{e2}}|^2|U_{e2}|^2+|\widetilde{U_{e3}}|^2|U_{e3}|^2),
  \label{eqn:solar prediction}
\end{equation}
where $|\widetilde{U_{ei}}|$ is an effective mixing matrix element where $^8B$ neutrinos are produced in the sun, and $|U_{ei}|$ is the mixing matrix element in vacuum. The effective mixing matrix elements depend on the charged-current potential at the neutrino source. The NC potential is not considered, as in this work active-sterile mixing angles ($\theta_{A-S}$) are assumed to be tiny compared to the mixing angle in the active section ($\theta_{ij}$), \textit{i.e.} $\theta_{A-S}\ll\theta_{ij}$. We consider $^8B$ neutrinos generated at a single point where the matter density is 93.11~$\mathrm{g/cm^3}$ and with the energy $6.44$~MeV \cite{Bahcall:2004pz}. $\Phi_{^8B}$ is the predicted $^8B$ solar neutrino flux from solar model BS05(AGS,OP) \cite{Bahcall:2004pz}, which is $4.51\times10^{6}~\mathrm{cm^{-2}s^{-1}}$. Simultaneously fitting results from all three SNO operational phases, the $\chi^2_{solar}$ is defined as
\begin{equation}
  \chi^{2}_{solar} = \sum\limits^{\mathrm{SNO~phase}}\frac{\left[\Phi^{pred}_{CC}-\Phi^{obs}_{CC}\right]^2}{\sigma_{CC}^{2}}.
  \label{eqn:solar chi2}
\end{equation}

\subsection{Long-baseline accelerator experiments}

Latest data from NOvA~\cite{Acero:2019ksn} and T2K~\cite{Abe:2018wpn, Abe:2019vii} from $\nu_{\mu}(\bar{\nu_{\mu}})$ disappearance and $\nu_{e}(\bar{\nu_{e}})$ appearance channels, for both neutrino and anti-neutrino modes are included in this analysis. The NOvA data consist of an exposure of $0.89\times10^{21}$~POT neutrino and $1.23\times10^{21}$ POT anti-neutrino beams. The T2K data include measurements from a neutrino beam mode with an exposure of $1.49\times10^{21}$ POT, and an anti-neutrino beam with an exposure of $1.64\times10^{21}$ POT. Each experiment collects several data samples, $\eg$ the T2K experiment has $\nu_{\mu}(\bar{\nu}_{\mu})$-enriched samples, $\nu_{e}(\bar{\nu}_{e})$-enriched samples, and $\nu_e$~CC1$\pi^{+}$ sample. We take the spectra sampled by far detectors.
\begin{table}[h!]
  \caption{Summary table of data samples of long-baseline accelerator experiments.}
  \begin{center}
  \begin{spacing}{1.2}
  \begin{tabular}{c|c|c|c}
    \toprule
    Exp			&  Sample $\alpha$					& Signal events $\eta$	&  Background $bkg$  \\
    \hline\hline
    \multirow{5}*{T2K}	& \multirow{2}*{$\nu_{\mu}$ disappearance}	& $(\nu_{\mu}+\bar{\nu}_{\mu})$CCQE,	&  \multirow{2}*{NC}  \\
    ~					& ~											& $(\nu_{\mu}+\bar{\nu}_{\mu})$CCnonQE & ~ \\
    \cline{2-4}
    ~                  	& \multirow{2}*{$\bar{\nu}_{\mu}$ disappearance}	& $(\nu_{\mu}+\bar{\nu}_{\mu})$CCQE,	&  \multirow{2}*{NC}  \\
    ~					& ~												& $(\nu_{\mu}+\bar{\nu}_{\mu})$CCnonQE & ~ \\
    \cline{2-4}
    ~                  	& $\nu_{e}$ appearance	& $(\nu_e+\bar{\nu}_e)$CCQE	&  NC, Beam~$(\nu_e+\bar{\nu}_e)$ \\
    \cline{2-4}
    ~                  	& $\bar{\nu}_{e}$ appearance	& $(\nu_e+\bar{\nu}_e)$CCQE	&  NC, Beam~$(\nu_e+\bar{\nu}_e)$ \\
    \cline{2-4}
    ~                  	& $\nu_e$ CC1$\pi^{+}$ appearance	& $\nu_e$CC1$\pi^{+}$	&  NC, Beam~$(\nu_e+\bar{\nu}_e)$ \\
    \hline
    \multirow{4}*{NOvA}	& $\nu_{\mu}$ disappearance    		& $(\nu_{\mu}+\bar{\nu}_{\mu})$CCQE	&  NC, Cosmic \\
    \cline{2-4}
    ~					& $\bar{\nu}_{\mu}$ disappearance	& $(\nu_{\mu}+\bar{\nu}_{\mu})$CCQE	&  NC, Cosmic \\
    \cline{2-4}
    ~                  	& $\nu_{e}$ appearance            	& $(\nu_e+\bar{\nu}_e)$CCQE	&  NC, Cosmic \\
    \cline{2-4}
    ~                  	& $\bar{\nu}_{e}$ appearance       	& $(\nu_e+\bar{\nu}_e)$CCQE	&  NC, Cosmic \\
    \bottomrule
  \end{tabular}
  \end{spacing}
  \label{tab:LBN_data}
  \end{center}
\end{table}

To analyses the published results, we refer to Ref.~\cite{Esteban:2018azc} and predict the spectra according to available information, including neutrino flux spectra, cross-sections, energy responses, and backgrounds. We summarise the data samples used in this analysis in Table.~\ref{tab:LBN_data}, along with the signal and background components we analysed. For a $\nu_{\alpha}$-enriched data sample $\alpha$, the estimated number of event $N_{i,\alpha}$ in the $i$th energy bin is given by 
\begin{equation}
\begin{aligned}
  N^{\alpha}_{i} =& N^{\alpha}_{bkg,i} + (1+\theta_{\alpha})\int^{E_{i+1}}_{E_{i}} dE_{rec} \int^{\infty}_{0} dE_{\nu} \sum^{\eta} R_{\eta}(E_{rec},E_{\nu}) \\
  & \times \frac{d\Phi}{dE_{\nu}} \sigma_{\eta}(E_{\nu}) \epsilon_{\eta}(E_{\nu}) P^{NU}_{\nu_{\mu}(\bar{\nu}_{\mu})\rightarrow\nu_{\alpha}}(E_{\nu}),
\end{aligned}
  \label{eqn:accelerator prediction}
\end{equation}
where $N^{\alpha}_{bkg,i}$ is the number of background events in the energy bin $i$, which we have extracted from NOvA~\cite{Acero:2019ksn} and T2K~\cite{Tsui:2018sterile, Vlad:2019Pred} predicted spectra. We do not include any oscillations for the background components. $E_{\nu}$ and $E_{rec}$ are true and reconstructed neutrino energy, respectively. $\epsilon_{\eta}(E_{\nu})$, as a function of $E_\nu$, is the detection efficiency for the event $\eta$.  $[E_{i}, E_{i+1}]$ are bin edges. $\frac{d\Phi}{dE_{\nu}}$ is the incident $\nu_{\mu}(\bar{\nu_{\mu}})$ flux. We extract the neutrino flux spectra for T2K from Ref.~\cite{Izmaylov:2017rec}, and for NOvA from Ref.~\cite{Sanchez:2018nova}. The flux spectra are rescaled by the number of protons on target and the fiducial number of nucleons in the detector. $R_{\eta}(E_{rec},E_{\nu})$ is the energy response function given by:
\begin{equation}
  R_{\eta}(E_{rec},E_{\nu}) = \frac{1}{\sqrt[2]{2\pi}\delta E_{\nu}}\exp\left( -\frac{(E_{\nu}-E_{rec})^2}{2(\delta E_{\nu})^2} \right)
\end{equation}
and $\eta$ denotes different interactions \textit{such as} CCQE, CC1$\pi^+$, or CC-nonQE. The NOvA far detector energy resolution $\delta E_{\nu}/E_{\nu}$ is $9\%$ for $\nu_{\mu}(\bar{\nu}_{\mu})$ CCQE events, and $11\%$ for $\nu_{e}(\bar{\nu}_{e})$ CCQE events\cite{Baird:2018rec}. The energy resolution for the T2K far detector is described in the formula,
\begin{equation}
  \delta E_{\nu}/E_{\nu} = \sqrt{a^2+\frac{b^2}{E_{\nu}}+\frac{c^2}{E_{\nu}}},
\end{equation}
which fits to $14\%$ at $0.7$~GeV, $8.4\%$ at $1$~GeV, and $5\%$ at $2$~GeV taken from Refs.~\cite{Hagiwara:2009bb, Abe:2014ugx}. Concerning the energy loss as hadron energy may not be properly measured in T2K far detector, we assume a $-0.4$~GeV energy shift for CC non-QE components. $\sigma_{\eta}(E_{\nu})$ is the cross-section for interaction $\eta$. We use the default cross section generated by GENIE~\cite{Andreopoulos:2009rq} for NOvA, and extracted cross-section for T2K from Ref.~\cite{Izmaylov:2017rec}. $P_{\nu_{\mu}\rightarrow\nu_{\alpha}}$ is the $\nu_{\mu}\rightarrow\nu_{\alpha}$ oscillation probability, taking into account matter effects for neutrino passing through the Earth's crust. However, only charged-current potential is taken into account. Similar to solar neutrino experiments, we assume that these NC matter effects from the active-sterile mixing are negligible, as these mixing angles are assumed to be tiny in this work. Including NC matter potentials, the neutrino oscillation patter will depend on the number of sterile neutrinos, which is set to be unknown in this analysis. This assumption can be realised, and we leave it to the future work.

We also take into consideration 5 $\nu_{\tau}$ events observed in a $\nu_{\mu}$ beam in OPERA. A rate-only fit to the observed number of events is performed. The prediction for the $\nu_\tau$ event number in OPERA is expanded
\begin{equation}
\begin{aligned}
  N &= N_{bkg} + \epsilon \int^{E_{max}}_{E_{min}} dE_{\nu} \frac{dN^{no-osc}(\theta_{best})}{dE_{\nu}} P^{NU}_{\nu_{\mu}\rightarrow\nu_{\tau}}(\theta,E_{\nu})
  \\
  &= N_{bkg} + \epsilon \int^{E_{max}}_{E_{min}} dE_{\nu} \frac{dN^{osc}(\theta_{best})}{P^{3\nu}_{\nu_{\mu}\rightarrow\nu_{\tau}}(\theta_{best},E_{\nu})dE_{\nu}} P^{NU}_{\nu_{\mu}\rightarrow\nu_{\tau}}(\theta,E_{\nu})  , 
\end{aligned}
\label{eqn:OPERA prediction}
\end{equation}
where the no-oscillation prediction $\frac{dN^{\nu_{\tau}}_{no-osc}(\theta_{best})}{dE_{\nu}}$ is obtained from the best-fit prediction $\frac{dN^{\nu_{\tau}}_{osc}(\theta_{best})}{dE_{\nu}}$~\cite{Serhan:2015Obs}, divided by the best-fit oscillation probability $P^{3\nu}_{\nu_{\mu}\rightarrow\nu_{\tau}}(\theta_{best},E_{\nu})$. The estimated number of background $N_{bkg}$~\cite{Agafonova:2015jxn} is 0.25 and is assumed to be unchanged.

Eventually, the estimated spectra are compared to the observed spectra $N^{obs}_{i}$ (total number of $\nu_\tau$ events in the case of OPERA) by constructing a Poissonian $\chi^2_{acc}$, 
\begin{equation}
  \chi^2_{acc} = \sum_{i}2\left(N^{pred}_{i}-N^{obs}_{i}+N^{obs}_{i}\log\left(\frac{N^{obs}_{i}}{N^{pred}_{i}}\right)\right) + \sum_{j}\frac{\theta_{j}^2}{\sigma_{j}^2},
  \label{eqn:accelerator chi2}
\end{equation}
where $\sigma_{j}$ are uncertainties~\cite{Esteban:2018azc} of nuisance parameters $\theta_{x}$ and their values are summarized in Table.\ref{tab:Systematic_summary}. The normalization factors for NOvA $\nu_{\mu}$ disappearance channel and $\bar{\nu}_{\mu}$ disappearance channel are fully correlated. In other words, these two channels share a common nuisance parameter.
\begin{table}[h!]
  \caption{Summary table of systematic uncertainties. Normalization factors are applied to the following measurements}
  \begin{center}
  \begin{tabular}{c|c|c}
    \toprule
    Experiment  &  Sample  &  $\sigma_{j}$  \\
    \hline\hline
    \multirow{5}*{T2K}  & $\nu_{\mu}$ disappearance        &  3\%  \\
    ~                   & $\bar{\nu}_{\mu}$ disappearance  &  4\%  \\
    ~                   & $\nu_{e}$ appearance             &  4.7\%  \\
    ~                   & $\bar{\nu}_{e}$ appearance       &  5.9\%  \\
    ~                   & $\nu_e$ CC1$\pi^{+}$ appearance          &  14.3\%  \\
    \hline
    \multirow{3}*{NOvA} & $\nu_{\mu}(\bar{\nu}_{\mu})$ disappearance    &   6\%  \\
    ~                   & $\nu_{e}$ appearance             &  5\%  \\
    ~                   & $\bar{\nu}_{e}$ appearance       &  6\%  \\
    \bottomrule
  \end{tabular}
  \label{tab:Systematic_summary}
  \end{center}
\end{table}

It is worthy to note that in analyses for long-baseline accelerator experiments, some of the inputs are derived assuming unitarity. For example, cross-sections for the T2K experiments were tuned according to T2K near detector data. In presence of sterile neutrinos, the true oscillation probability deviates from that assuming unitarity. This can be a potential improvement into our future work.

\subsection{Limits to sterile neutrinos}
 
Results from sterile neutrino searches are used in this analysis to provide extra information of the sterile sector.
Global fits with data from various experiments \cite{Dentler:2018sju, Giunti:2019aiy} report exclusion limits on the sterile neutrino mixing. We treat the exclusion limits as constraints to the non-unitarity. The constraints are considered to be Gaussian, with central values being zero. 
We note for large mass-squared differences, the sterile neutrino mixing driven oscillations are averaged out, and the exclusion limits on sterile neutrino mixing parameters will be independent from mass-squared differences. Proper choices are made with exclusion limits for $\Delta m^2_{41} \geq 0.1 \mathrm{eV}^2$ \cite{Fong:2016yyh}. 

Constraint provided by Ref.~\cite{Dentler:2018sju, Giunti:2019aiy} are for $|U_{\alpha4}|^2$ and $4|U_{e4}|^2|U_{\mu4}|^2$, which cannot be used directly in our analysis. Therefore, we need a translation.
Limits on $|U_{\alpha4}|^2$ are interpreted as limits on the variation from unity of the normalization of row $\alpha$, i.e. $1-\sum^{3}_{i=1}|U_{\alpha i}|^2~(\alpha=e,\mu,\tau)$. Limits on $4|U_{e4}|^2|U_{\mu4}|^2$ are interpreted as limits on \\
$4\left(1-\sum^{3}_{i=1}|U_{e i}|^2\right)\left(1-\sum^{3}_{i=1}|U_{\mu i}|^2\right)$. To constrain each of the non-unitarities, we define the $\chi^2_{sterile}$ as 
\begin{equation}
  \chi^2_{sterile} = \chi^2_{sterile, e} + \chi^2_{sterile, \mu} + \chi^2_{sterile, \tau} + \chi^2_{sterile, e\mu},
  \label{eq:chi_VSL}
\end{equation}
\begin{equation}
  \chi^2_{sterile, \alpha} = \frac{\left(1-\sum^{3}_{i=1}|U_{\alpha i}|^2\right)^2}{\sigma_{\alpha}^2},~~~~\mathrm{for}~ \alpha = e, \mu, \tau,
\end{equation}
\begin{equation}
  \chi^2_{sterile, e\mu} = \frac{\left[4\left(1-\sum^{3}_{i=1}|U_{e i}|^2\right)\left(1-\sum^{3}_{i=1}|U_{\mu i}|^2\right)\right]^2}{\sigma_{e\mu}^2},
\end{equation}
where $\sigma_{\alpha}$ are the exclusion limits at $1\sigma$ confidence level. Shown in Table.\ref{tab:Sterile Data Summary} are the exclusion limits and the non-unitarities they constrain. Exclusion limits reported at certain confidence levels are recast to limits at $1\sigma$ confidence level, assuming Gaussian distribution. 

\begin{table}[h!]
  \caption{Limits on sterile neutrino mixing from global fits. Searches for sterile neutrinos with null results\protect\footnotemark are used. Exclusion limits at $\Delta m^2_{41} \geq 0.1 \mathrm{eV}^2$ are chosen. }
  \begin{center}
  \begin{tabular}{c|c|c}
    \toprule
    Non-unitarity            						& Data & Limit ($1\sigma$)  \\
    \hline\hline
    \multirow{2}*{$1-\sum^{3}_{i=1}|U_{e i}|^2$}    & \multirow{2}*{SK+DC+IC} & \multirow{2}*{0.0589}   \\
     & & \\
    \hline
    \multirow{2}*{$4\left(1-\sum^{3}_{i=1}|U_{e i}|^2\right)\left(1-\sum^{3}_{i=1}|U_{\mu i}|^2\right)$}
     								& \multirow{2}*{OPERA($\nu_{e}$)} & \multirow{2}*{0.00713}  \\
     & & \\
    \hline
    \multirow{2}*{$1-\sum^{3}_{i=1}|U_{\mu i}|^2$}
                              & CDHSW+MNS+SB & \multirow{2}*{0.0061}  \\
                              & +MB+SK+DC+IC        & \\
    \hline
    \multirow{2}*{$1-\sum^{3}_{i=1}|U_{\tau i}|^2$}
                              & CDHS+MNS+NOvA & \multirow{2}*{0.0659}  \\
                              & +MB+SK+DC+IC+SNO    & \\
    \bottomrule
  \end{tabular}
  \end{center}
  \label{tab:Sterile Data Summary}
\end{table}
\footnotetext{Atmospheric neutrino data is from Super Kamiokande (SK), Deep Core (DC), and IceCube (IC). Accelerator charged current interaction data is from CDHSW, SciBooNE (SB), MiniBooNE (MB), MINOS$\&$MINOS+ (MNS), OPERA. Neutral current interaction data includes experimental data from SNO, NOvA, CDHS, MINOS$\&$MINOS+, Super Kamiokande, Deep Core, IceCube, and MiniBooNE.}

\subsection{$\chi^2$ function}

To combine all data from above experiments to constrain matrix elements of $U^{NU}$, we sum up all $\chi^2$ values from Eqs.~(\ref{eqn:reactor-mid total},~\ref{eqn:reactor-long chi2},~\ref{eqn:solar chi2},~\ref{eqn:accelerator chi2},~and \ref{eq:chi_VSL}) and define a total $\chi^2$ value,
\begin{equation}
  \chi^2_{total} (U^{NU}, \Delta m^2_{21}, \Delta m^2_{31}, \vec{\theta}) = \chi^2_{ALL~MR} + \chi^2_{LR} + \chi^2_{solar} + \chi^2_{ALL~acc}  + \chi^2_{sterile}.
\label{eqn:chi2}
\end{equation}
$\chi^2_{total}$ is a function of $U^{NU}$, two mass-squared differences $\Delta m_{21}^2$ and $\Delta m_{31}^2$ and a vector $\vec{\theta}$. In more details, inside $U^{NU}$ there are $13$ mixing parameters, including $9$ absolute values and $4$ phases. 
The vector of nuisance parameters $\vec{\theta}$ collects all normalization factors applied to the prediction for KamLAND Eq.~(\ref{eq:KamLAND_pred}) and accelerator neutrino experiments Eq.~(\ref{eqn:accelerator prediction}). For the ease of calculation, we fixed the mass-squared differences as $\Delta m^2_{21} = 2.51\times10^{-3} \mathrm{eV}^2$ and $\Delta m^2_{32} = 7.53\times10^{-5} \mathrm{eV^2}$, assuming normal mass ordering. Except for the constrained parameters and two mass-squared differences, the $\chi^2_{total}$ is minimized over all parameters by the TMinuit.Migrad minimizer~\cite{Brun:1997pa}. Finally, in our simulation, we impose the input hypothesis satisfying constraints such as Eq.~(\ref{eq:geometric_leq1}) and Cauchy-Schwarz inequalities Eq.~(\ref{eqn:Cauchy-Schwarz-Active_1})-(\ref{eqn:Cauchy-Schwarz-Active_2}).

\section{Results}\label{sec:result}

After combining all data from medium-baseline reactor, long-baseline reactor, solar, long-baseline accelerator experiments that are introduced in Sec.~\ref{sec:setting}, we will firstly present the resultant constraints on $U^{NU}$ with Eq.~(\ref{eqn:chi2}) in this section. Both $3\nu$ unitarity and non-unitarity assumptions are considered. We will also visit the CP violation in the case of $3\nu$ non-unitarity, before showing the goodness of fit of the current data to the $3\nu$ unitarity assumption.

\subsection{Contraint on the matrix element}\label{sec:result_element}

\begin{figure}[h!]
\centering 
\includegraphics[width=5.4cm]{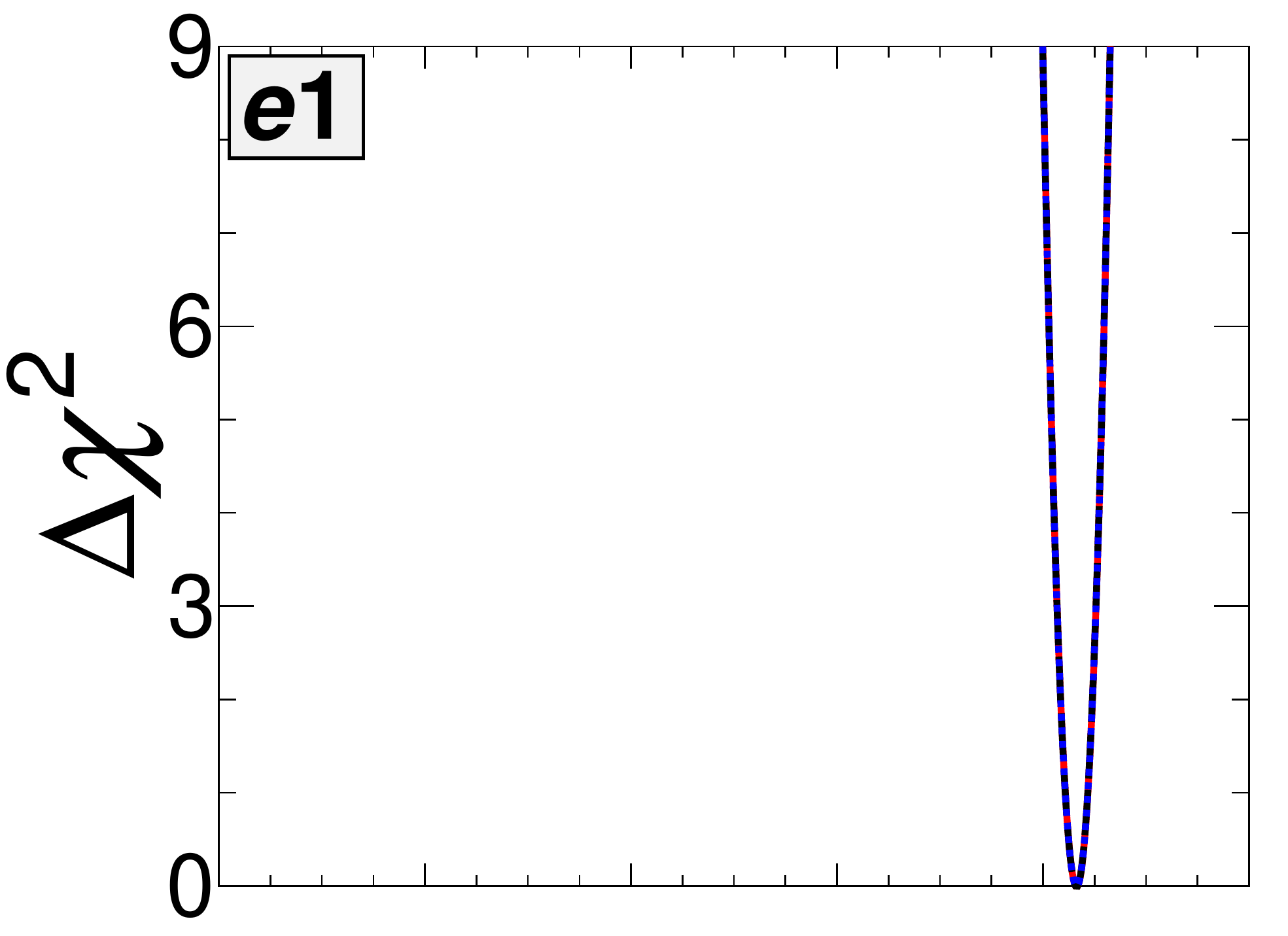}
\includegraphics[width=4.5cm]{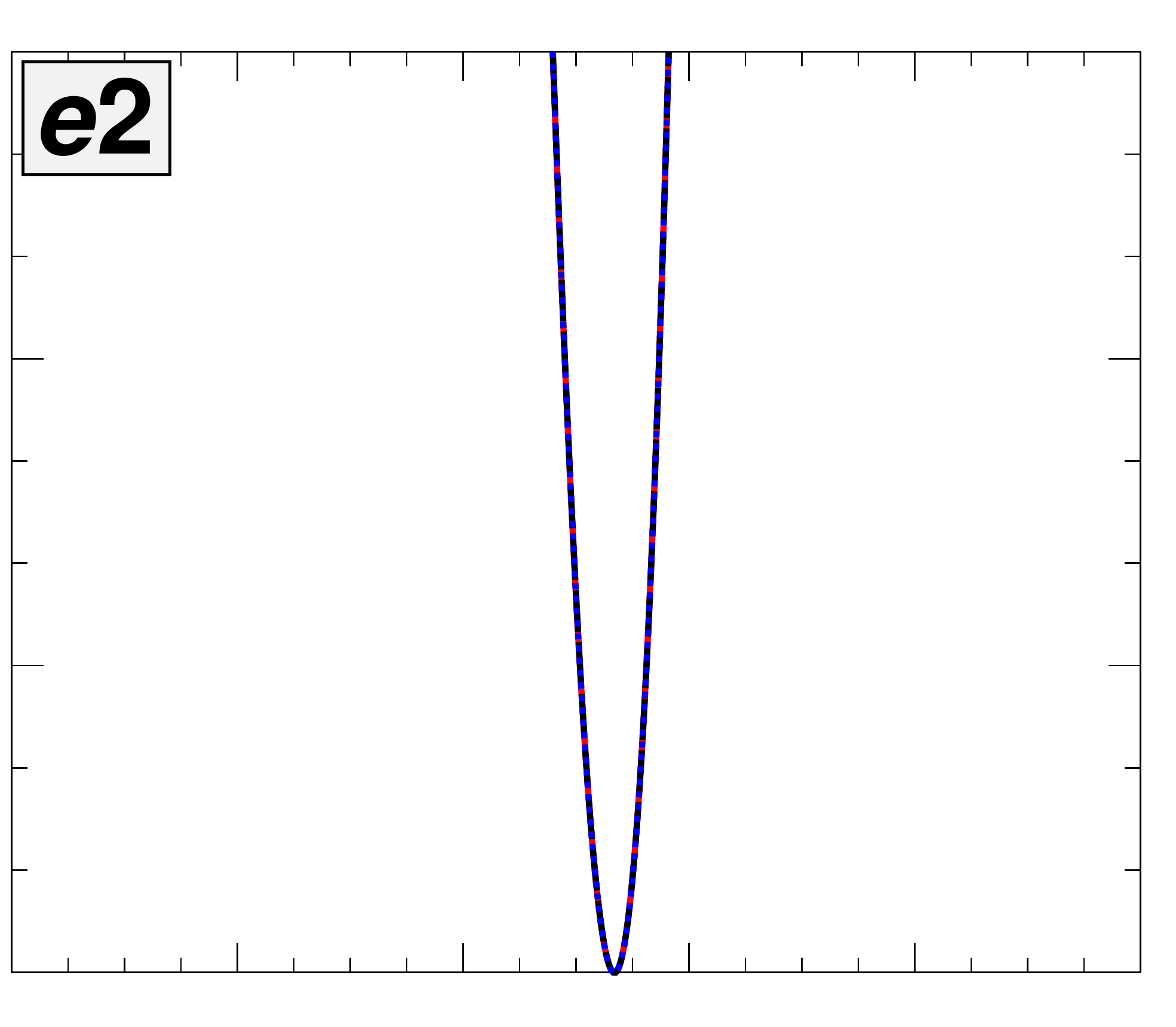}
\includegraphics[width=4.5cm]{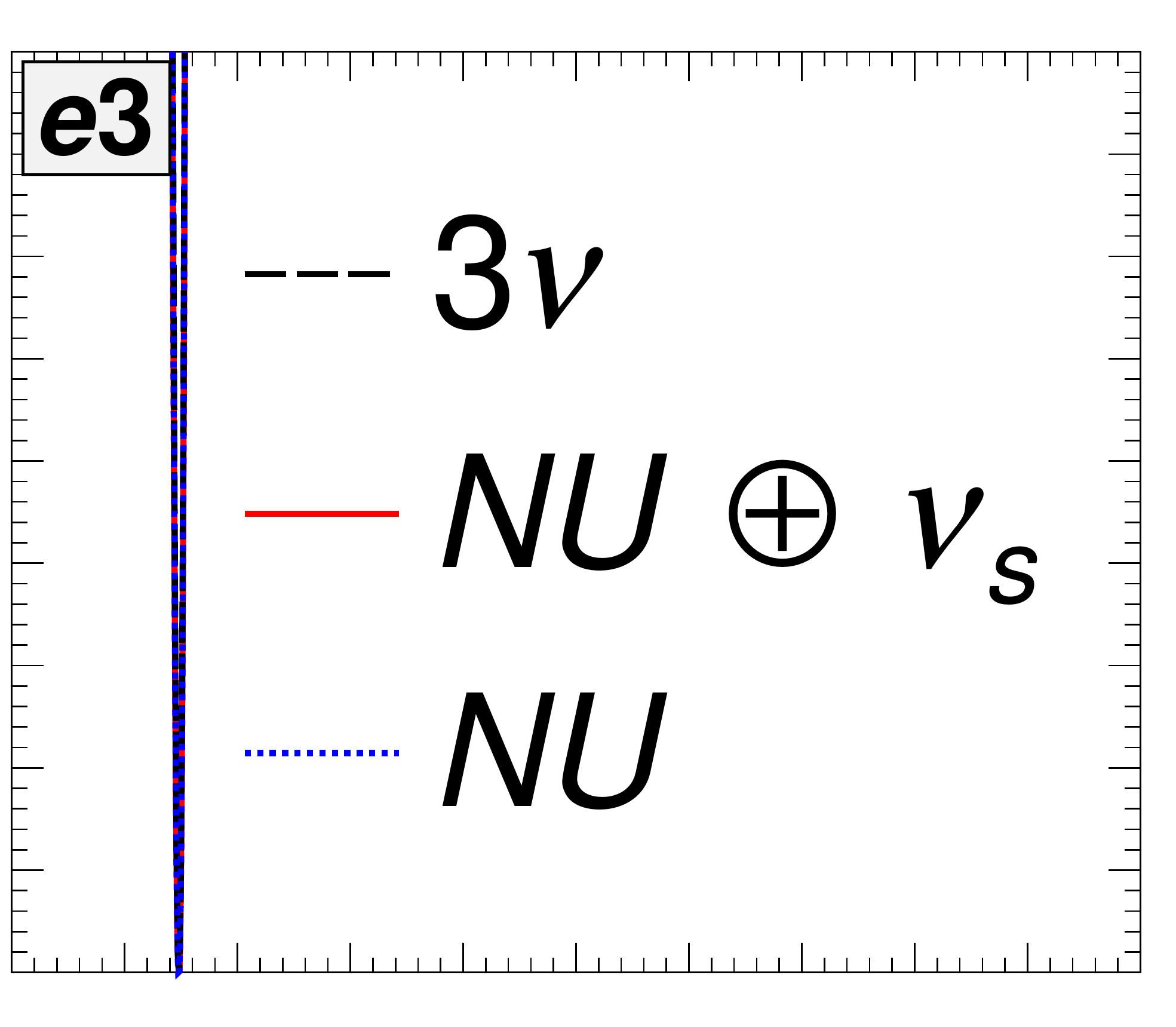}\\
\includegraphics[width=5.4cm]{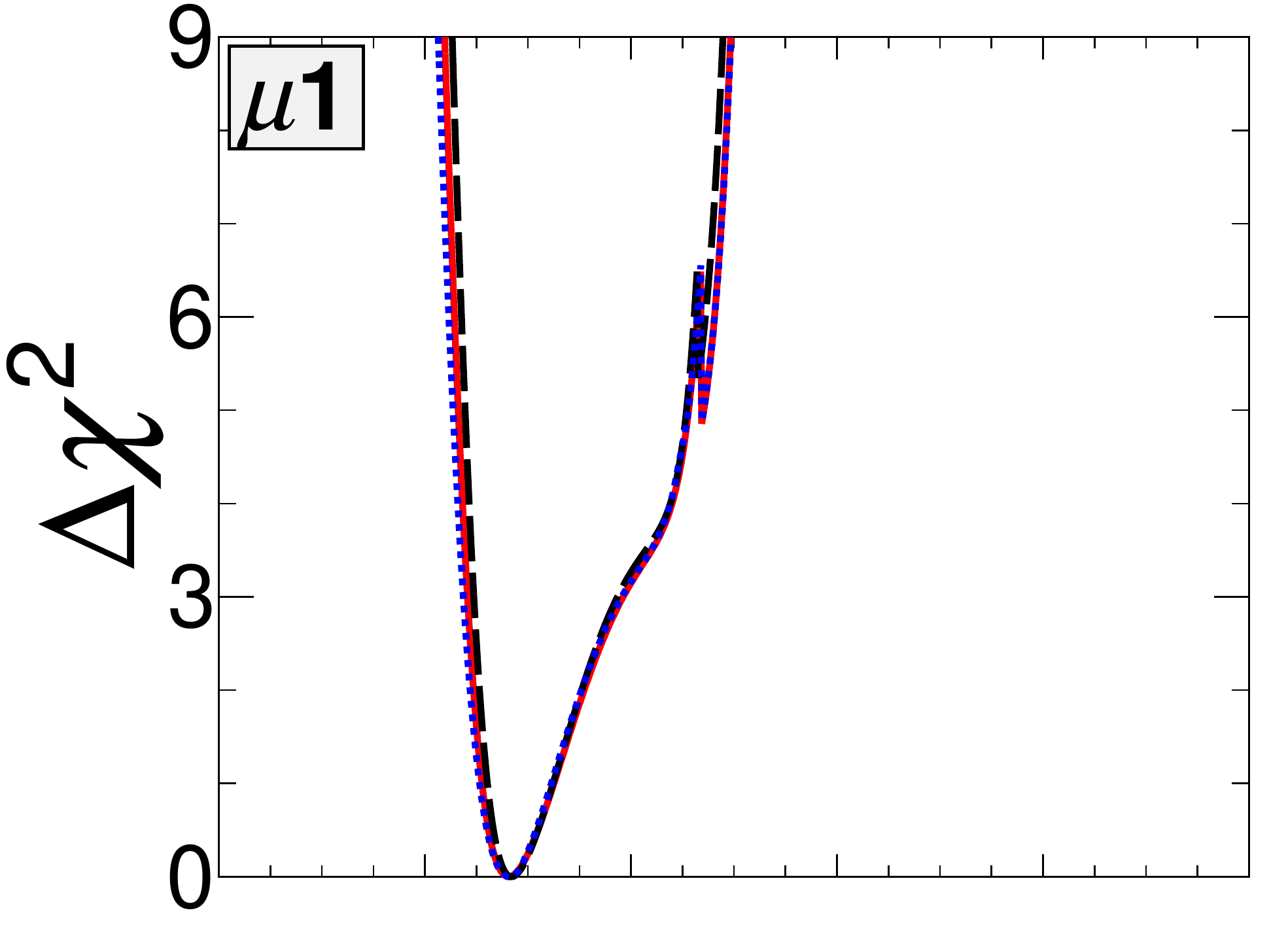}
\includegraphics[width=4.5cm]{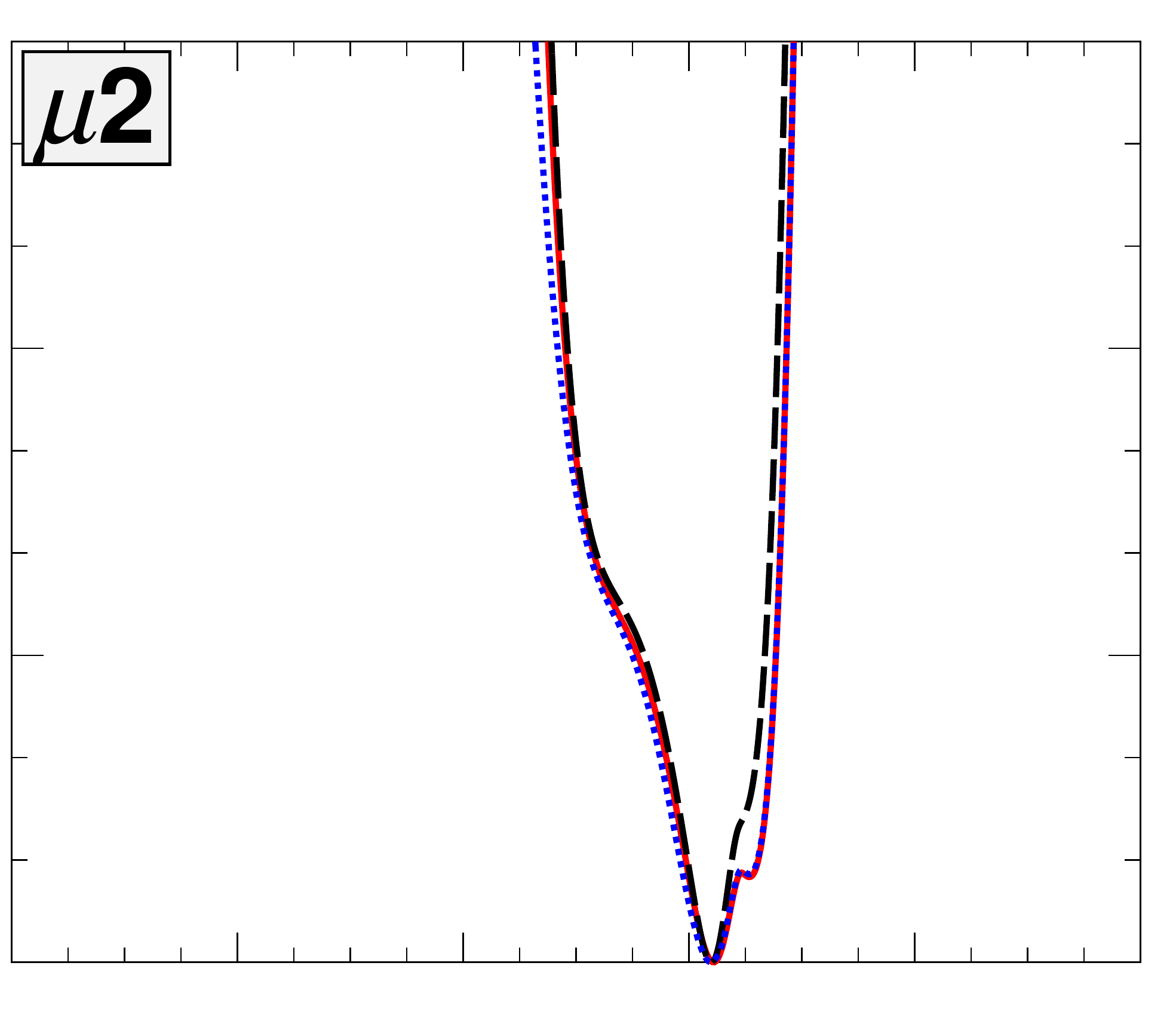}
\includegraphics[width=4.5cm]{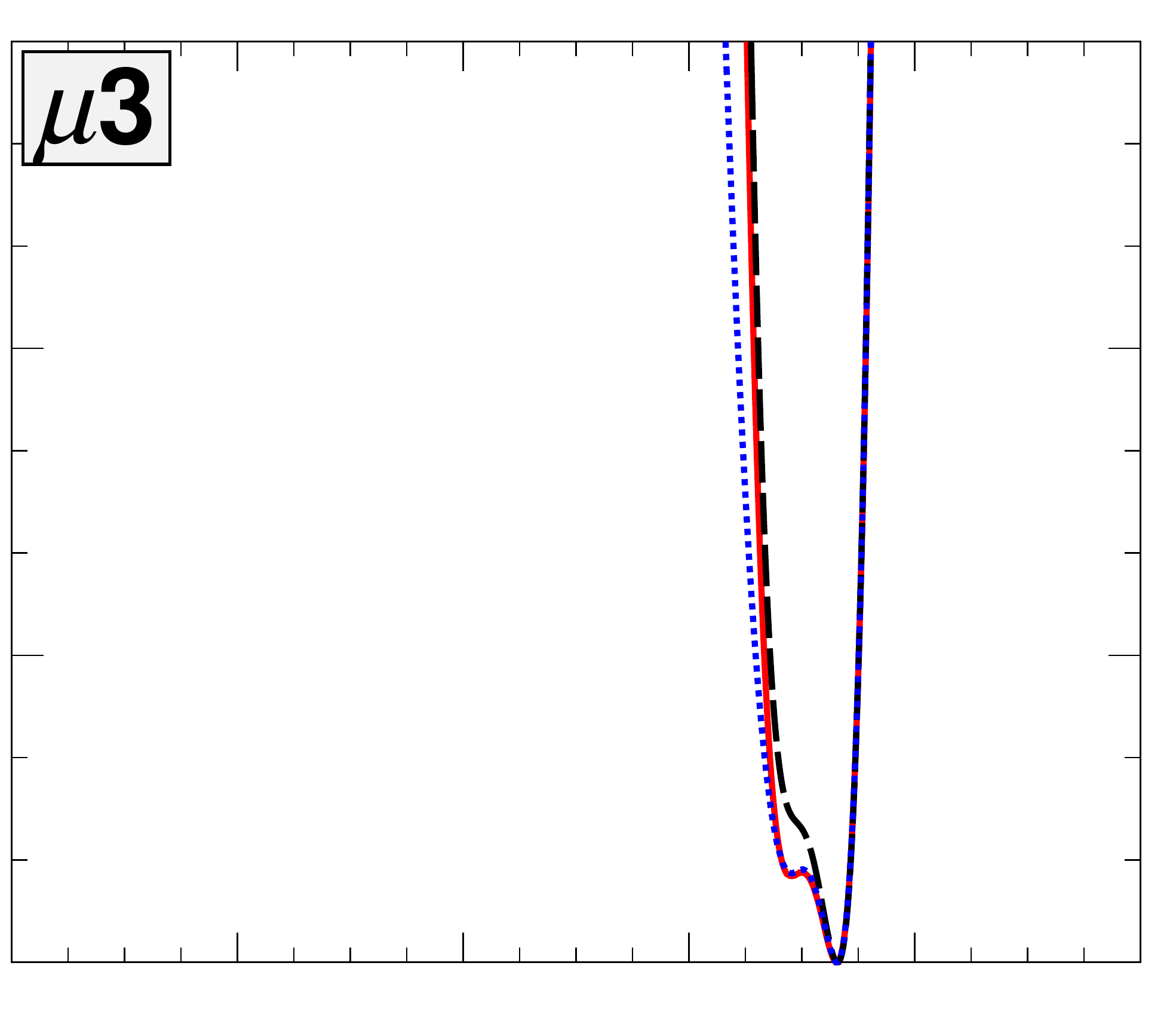}\\
\includegraphics[width=5.4cm]{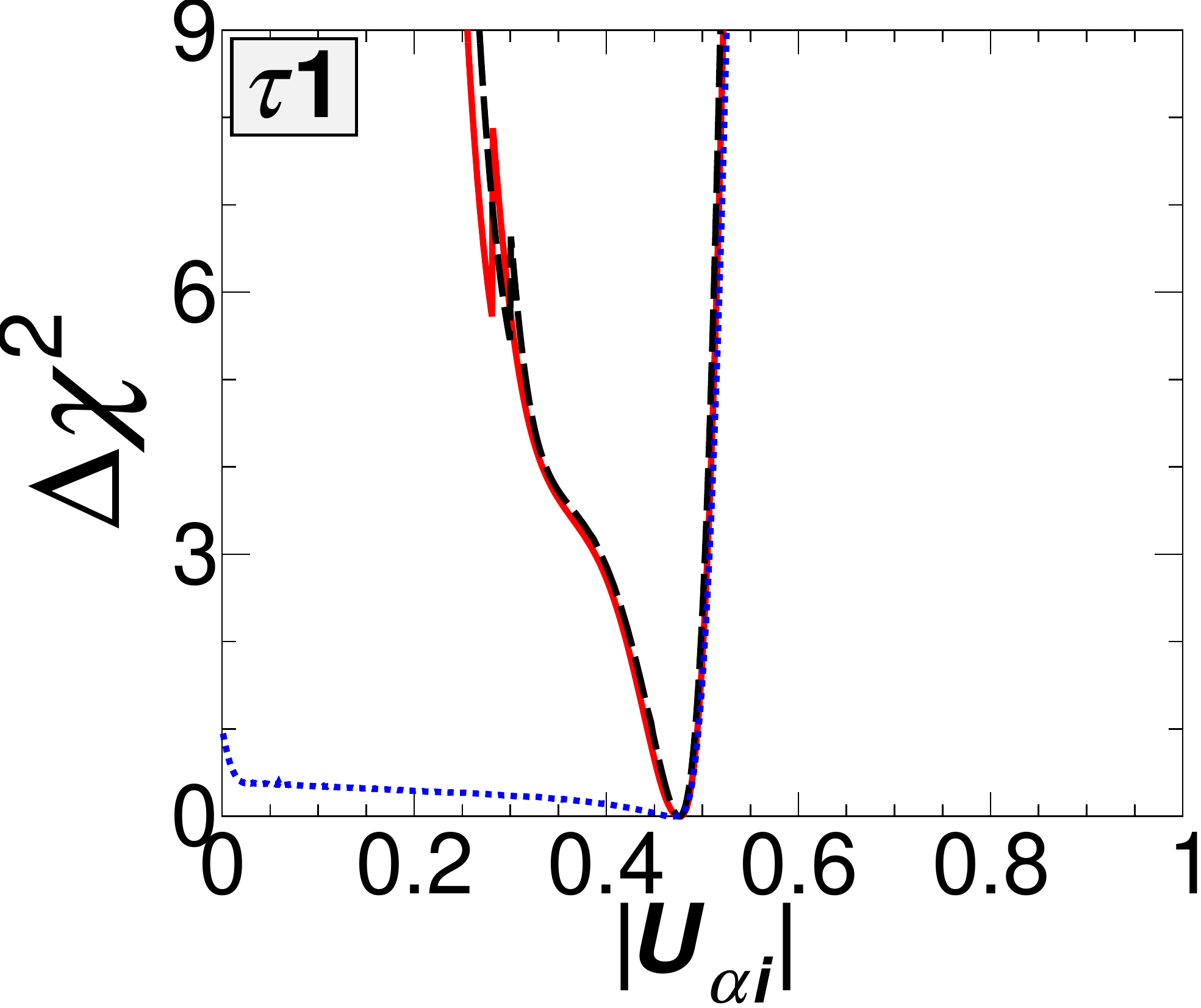}
\includegraphics[width=4.5cm]{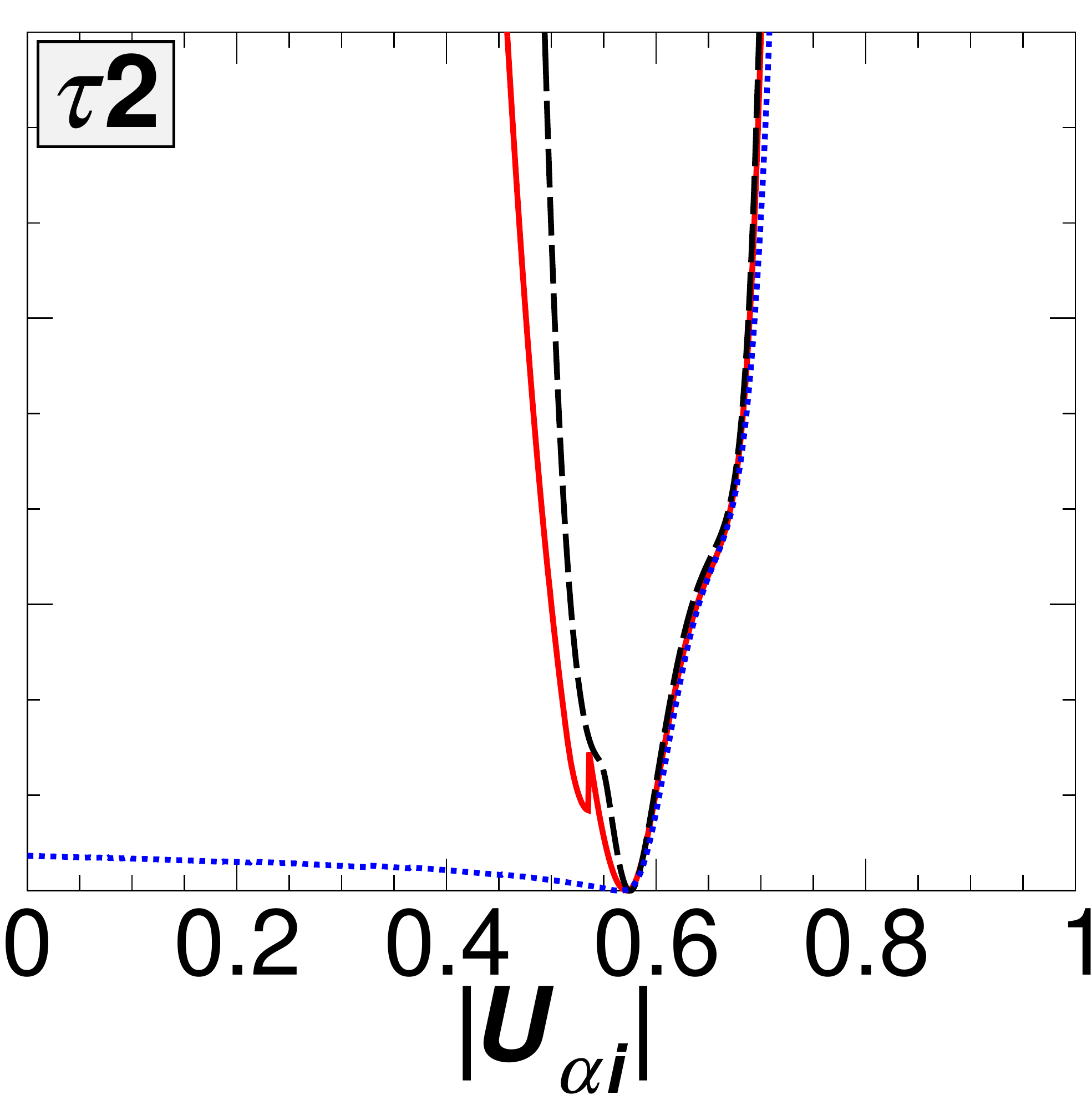}
\includegraphics[width=4.5cm]{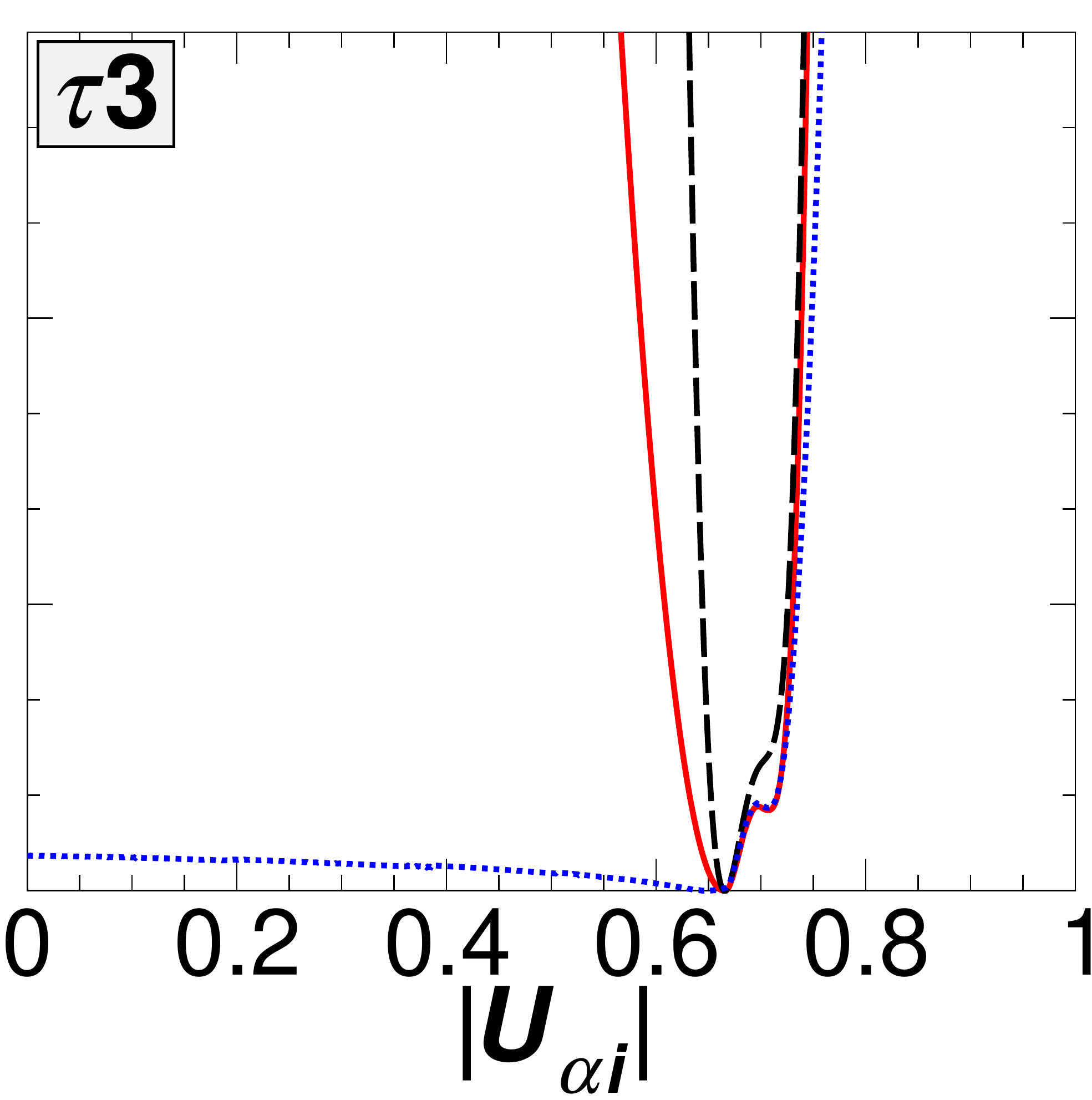}
\caption{The constraints on the moduli of the mixing matrix elements with (black dashed) and without (red solid) unitarity. The x-axes are the moduli of matrix elements, ranging from 0 to 1. The y-axes are the $\Delta\chi^2$, ranging from 0 to 9. A fit setting $\chi^2_{sterile}=0$, i.e. without using limits on sterile neutrinos, is also shown (blue dotted).}
\label{fig:constraint}
\end{figure}

We present the $\Delta\chi^2$ values against the absolute values of each matrix element in Fig.~\ref{fig:constraint} for different scenarios. 
We perform several groups of fittings with different assumptions: `$NU$' (blue-dotted) and `$NU\oplus\nu_s$' (red) are for the fit assuming non-unitarity with and without sterile search results, respectively; `$3\nu$' (black-dashed) is for the fit assuming $3\nu$ unitarity.

At the first glance of Fig.~\ref{fig:constraint}, the `$NU\oplus\nu_s$' fit is in good agreement with the `$3\nu$' fit. In addition, the $\Delta\chi^2$ behaves the same in all scenarios in the $\nu_{e}$ sector. This indicates that the current reactor and solar data seem to well match the unitarity of the $3\nu$ mixing matrix. 
From the narrowness of the 1-D $\Delta\chi^2$ curves, one sees that elements $|U_{ei}|$ have been precisely measured, with the $3\sigma$ errors being $0.065$, $0.103$, and $0.013$ for $|U_{e1}|$, $|U_{e2}|$, and $|U_{e3}|$, respectively. The precision measurements in the $\nu_{e}$ sector are passed to the $\nu_{\mu}$ sector via the Cauchy-Schwarz inequalities Eqs.~(\ref{eqn:Cauchy-Schwarz-Active_1}) and (\ref{eqn:Cauchy-Schwarz-Active_2}). 

\begin{figure}[h!]
\centering 
\includegraphics[width=7cm]{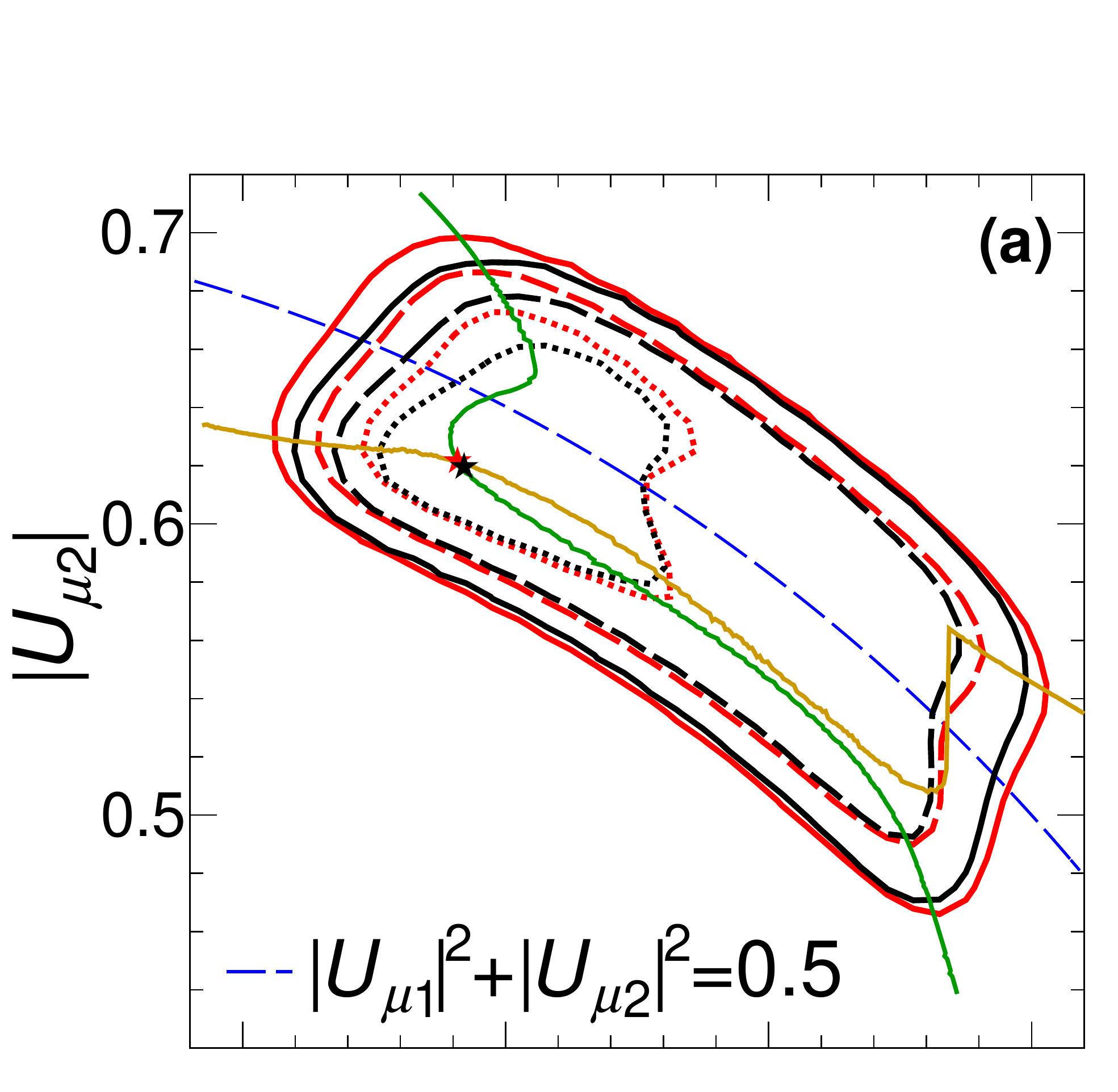}
\includegraphics[width=7cm]{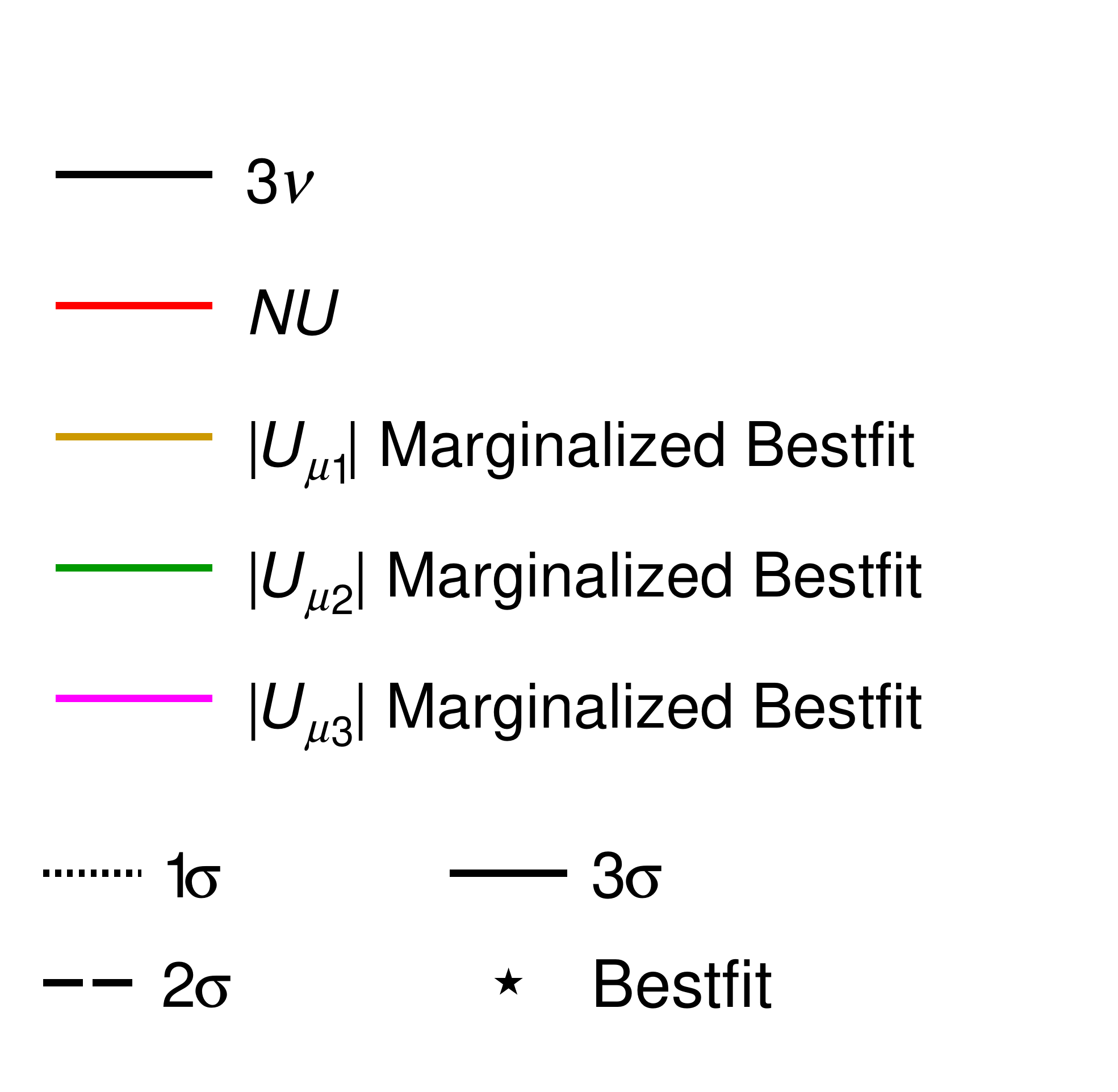}
\includegraphics[width=7cm]{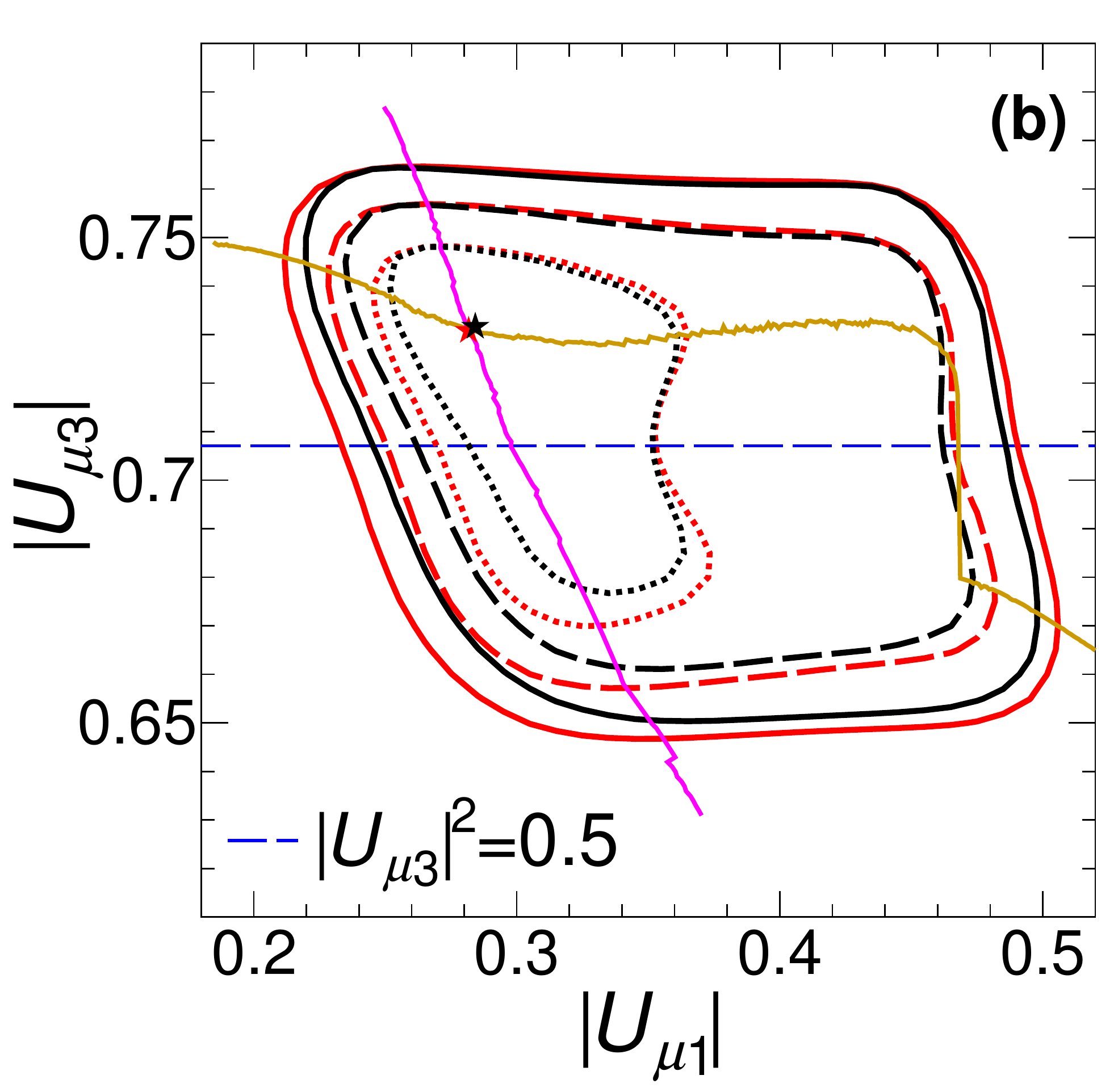}
\includegraphics[width=7cm]{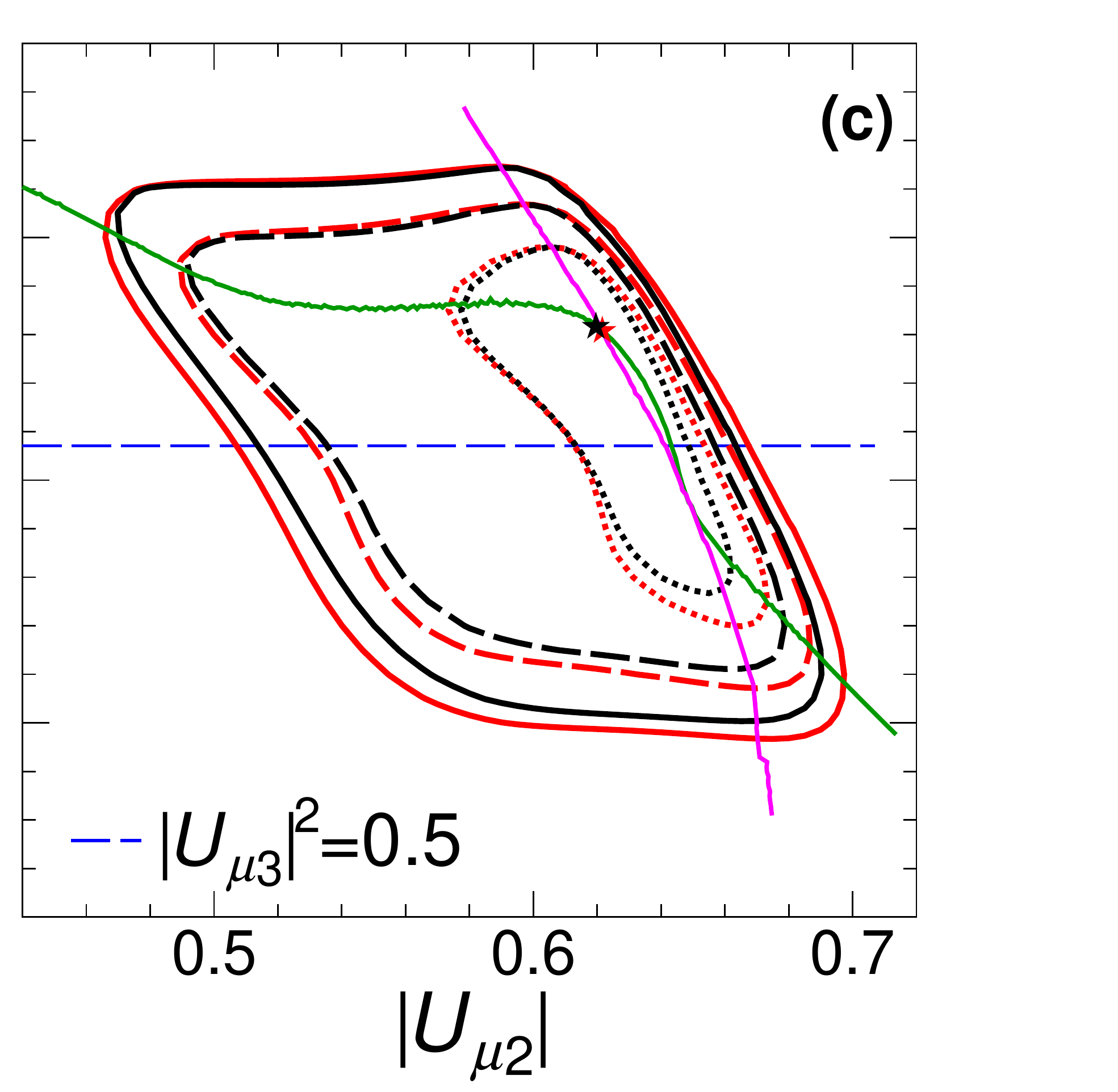}
\caption{The 2-D $\Delta\chi^2$ contours in $\nu_{\mu}$ sector. Allowed regions at 1$\sigma$ (dotted), 2$\sigma$ (dashed), 3$\sigma$ (solid) confidence level are shown, on (a) the $|U_{\mu1}|-|U_{\mu2}|$ plane, (b) the $|U_{\mu1}|-|U_{\mu3}|$ plane, and (c) the $|U_{\mu2}|-|U_{\mu3}|$ plane. Also shown are the bestfits of $|U_{\mu1}|$ (orange), $|U_{\mu2}|$ (cyan), and $|U_{\mu3}|$ (magenta), minimized over all of the other parameters, projecting on the 2-D planes. Auxillary lines (blue dashed) are drawn to help understanding the behaviours of the bestfits.} \label{fig:2D_mu}
\end{figure}

Let us draw readers attention to the spike-like structure in the 1-D $\Delta\chi^2$ curves at $|U_{\mu1}|\sim0.5$ and the dip structure around $|U_{\mu2}|\sim0.65$ and $|U_{\mu3}|\sim0.7$, in Fig.~\ref{fig:constraint}. In this analysis, $\nu_{\mu}\rightarrow\nu_{\mu}$ disappearance and $\nu_{\mu}\rightarrow\nu_{e}$ appearance measurements at NOvA and T2K provide predominant constraints to the $\nu_{\mu}$ sector. 
The leading terms of the oscillation probabilities in NOvA and T2K are easily derived from Eq.~(\ref{eqn:oscprob}), read
\begin{equation}
\begin{aligned}
  &-4\left(|U_{\mu 1}|^2+|U_{\mu 2}|^2\right)|U_{\mu 3}|^2,~~~~\mathrm{for}~\nu_{\mu}(\bar{\nu}_{\mu})\rightarrow\nu_{\mu}(\bar{\nu}_{\mu}),
  \\
  &-4\Re\left(U^{*}_{e1}U_{\mu1}+U^{*}_{e2}U_{\mu2}\right)U^{*}_{e3}U_{\mu3},~~~~\mathrm{for}~\nu_{\mu}(\bar{\nu}_{\mu})\rightarrow\nu_{e}(\bar{\nu}_{e}).
\end{aligned}
\label{eq:acc_leading}
\end{equation}
With the $3\nu$ unitarity assumption, the leading term for $\nu_{\mu}$ disappearance channel is $-4\left(1-|U_{\mu 3}|^2\right)|U_{\mu 3}|^2$, which is symmetric with respect to $|U_{\mu3}|=\frac{\sqrt{2}}{2}$, known as the octant degeneracy in the standard $3\nu$ scheme. This degeneracy can be disentangled by combining the $\nu_{\mu}$ disappearance measurements with the $\nu_{e}$ appearance measurements. Our unitary fit `$3\nu$' shows a preference for the upper \textit{octant}, which is consistent with the published results from NOvA and T2K, with the best-fit value $|U_{\mu3}|=0.733$. 
Without the $3\nu$ unitarity assumption, the same global best-fit is also found, while for `lower \textit{octant}'{\footnote{Here "\textit{octant}" means $|U_{\mu3}|=\frac{\sqrt{2}}{2}$, while upper(lower) \textit{octant} means $|U_{\mu3}|$ larger(smaller) than $\frac{\sqrt{2}}{2}$. Without the unitarity, the $3\nu$ mixing matrix can no longer be parameterized as 3 angles and 1 phase, and hence this octant-like degeneracy is denoted `\textit{octant}' in Italian font. }} $|U_{\mu3}|=0.689(\Delta\chi^2=0.84)$, and is consistent with $|U_{\mu3}|=\frac{\sqrt{2}}{2}$ within $1\sigma$. The spike-like structure around $|U_{\mu1}|=0.5$ and the dip structure around $|U_{\mu2}|=0.6$ are due the conversion of $|U_{\mu3}|$ from `upper \textit{octant}' to `lower \textit{octant}', left-to-right. More details will be introduced in Fig.~\ref{fig:2D_mu}.

We summarise our result as follows.
The best-fit points are the same for both non-unitary and unitary cases, 
\begin{equation}\label{eq:bestfit-non-unitary}
|U|^{NU}_{b.f.} = 
  \begin{pmatrix}
    0.832~~ & 0.535~~ & 0.148 \\
    0.281~~ & 0.622~~ & 0.730 \\
    0.477~~ & 0.572~~ & 0.664 
  \end{pmatrix}.
\end{equation}
The $3\sigma$ confidence intervals for the two fits are shown in Eq.~(\ref{eq:3sigma-non-unitary}) and Eq.~(\ref{eq:3sigma-unitary}), respectively. The results assuming unitarity are in a good consistence with the results of three-flavour global fit~\cite{1809173},
\begin{equation}\label{eq:3sigma-non-unitary}
|U|_{3\sigma}^{NU} = 
  \begin{pmatrix}
    0.800\rightarrow0.865~~ & 0.479\rightarrow0.582~~ & 0.141\rightarrow0.154 \\
    0.219\rightarrow0.497~~ & 0.475\rightarrow0.693~~ & 0.651\rightarrow0.761 \\
    0.255\rightarrow0.521~~ & 0.458\rightarrow0.700~~ & 0.566\rightarrow0.744 
  \end{pmatrix},
\end{equation}

\begin{equation}\label{eq:3sigma-unitary}
|U|_{3\sigma}^{3\nu} = 
  \begin{pmatrix}
    0.800\rightarrow0.865~~ & 0.480\rightarrow0.582~~ & 0.143\rightarrow0.153 \\
    0.226\rightarrow0.489~~ & 0.479\rightarrow0.685~~ & 0.655\rightarrow0.761 \\
    0.268\rightarrow0.518~~ & 0.493\rightarrow0.698~~ & 0.631\rightarrow0.740 
  \end{pmatrix}.
\end{equation}

We intend to investigate the correlations between the $3\nu$ mixing matrix elements when the conditions of $3\nu$ unitarity are not imposed. In Fig.~\ref{fig:2D_mu}, we plot $1$ to $3\sigma$ allowed regions on the $|U_{\mu i}|-|U_{\mu j}|$ plane, with (black) and without unitarity (red). On the $|U_{\mu 1}|-|U_{\mu 2}|$ plane, the allowed regions for both cases are expanded from left-up to right-down direction, indicating a negative correlation. 
This correlation is explained, if one looks at Eq.~(\ref{eq:acc_leading}), by noticing the $\nu_{\mu}$ disappearance measurements at NOvA and T2K only determine the combination of $|U\mu1|^2+|U_{\mu2}|^2$. 
By including $\nu_e$ appearance measurements in the fit, one finds the values of $U_{\mu1}e^{i(\phi_{e1}-\phi_{\mu1})}$ and $U_{\mu2}e^{i(\phi_{e2}-\phi_{\mu2})}$, given $|U_{e1}|$ and $|U_{e2}|$ known from reactor and solar neutrino experiments. Similar behaviours of the $|U_{\mu1}|-|U_{\mu3}|$ and $|U_{\mu2}|-|U_{\mu3}|$ contours, are caused by the same reason. We notice that even releasing the $3\nu$ unitarity restriction, the correlations do not change significantly. 

In addition to the correlations, the degeneracies appeared in Fig.~\ref{fig:constraint} are also shown in $|U_{\mu i}|-|U_{\mu j}|$ 2-D plots in Fig.~\ref{fig:2D_mu}. We plot the curves when $|U_{\mu3}|^2=0.5$ or $|U_{\mu 1}|^2+|U_{\mu 2}|^2=0.5$ in the case of unitarity, to show the maximal mixing. All of the contours are depressed around the maximal mixing. We also plot the local minimum (light blue and orange curves) with fixed $U_{\mu i}$ value, and find the discontinuity in Fig.~\ref{fig:constraint}, happens at the place where the local minimum switches the \textit{octant}.

\subsection{CP violation}

\begin{figure}[h!]
\centering
\includegraphics[width=7cm]{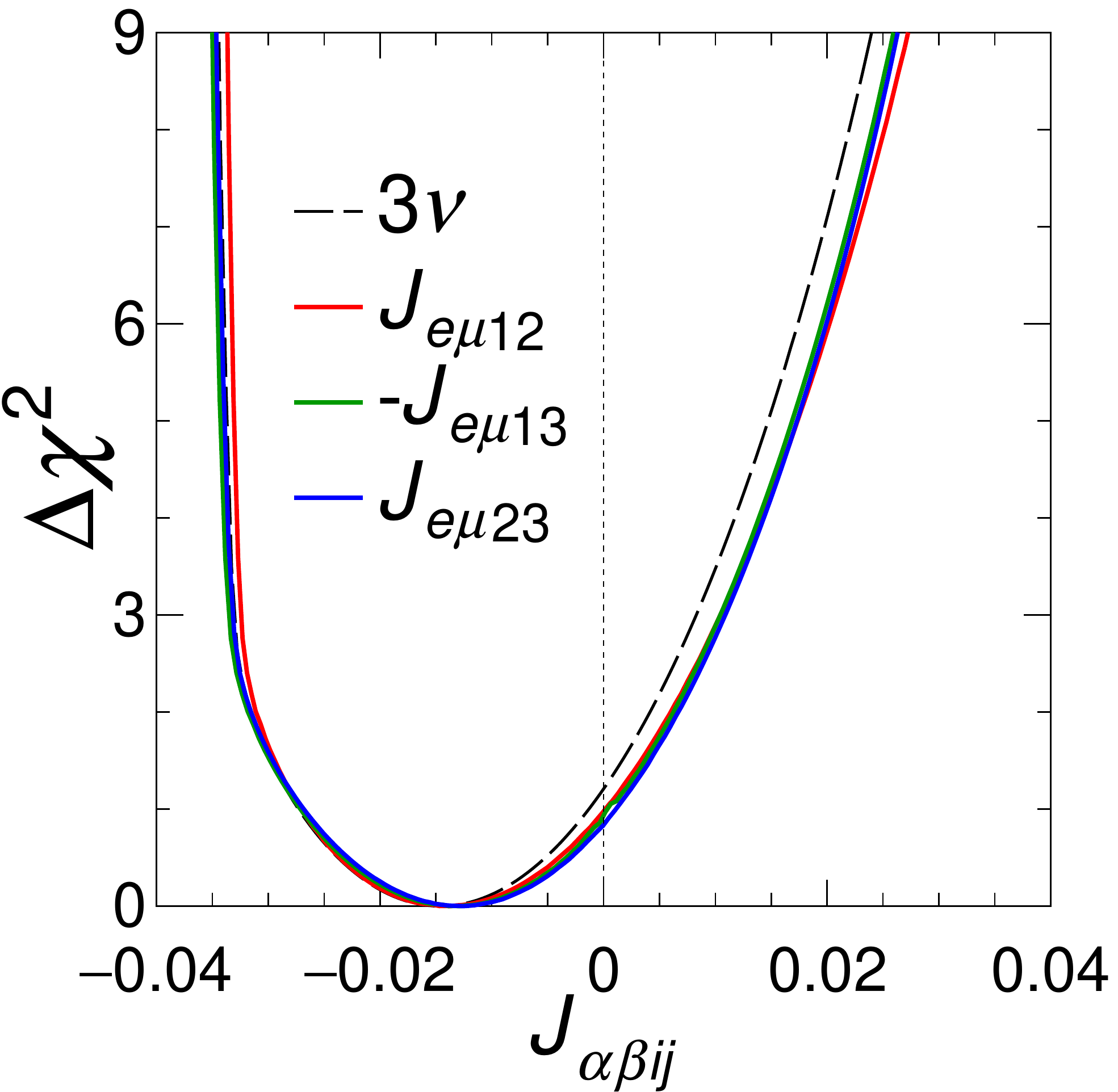}
\caption{1-D $\Delta\chi^2$ of the three Jarlskog factors associated to the $\nu_e-\nu_{\mu}$ sector: $J_{e\mu12} \equiv \Im[U_{e1}U_{\mu2}U^{*}_{e2}U^{*}_{\mu1}]$~(red), $J_{e\mu23} \equiv \Im[U_{e2}U_{\mu3}U^{*}_{e3}U^{*}_{\mu2}]$~(blue) and $J_{e\mu13} \equiv \Im[U_{e1}U_{\mu3}U^{*}_{e3}U^{*}_{\mu1}]$~(green). A minus sign is appended to $J_{e\mu13}$. Also shown is the $3\nu$ Jarlskog invariant~(black dashed). }
\label{fig:Jarlskog}
\end{figure}

The CP violation in the case of non-unitarity is currently measured in the form of Jarlskog factors $J_{\alpha\beta ij}$. For $U^{3\nu}$, an invariant Jarlskog factor characterizes CP violation, as expressed in Eq.~(\ref{eq:jarlskog inv}). Without the $3\nu$ unitarity assumption, the Jarlskog factor is no longer an invariant. Taking different combinations of $\alpha\beta ij$, there can be at most nine different Jarlskog factors. As we have little direct measurements of the $\nu_{\tau}$ sector, six of the Jarlskog factors related to the $\nu_{\tau}$ sector are not analysed. We show in Fig.~\ref{fig:Jarlskog} three Jarlskog factors associated to the $\nu_e-\nu_{\mu}$ sector: $J_{e\mu12} \equiv \Im[U_{e1}U_{\mu2}U^{*}_{e2}U^{*}_{\mu1}]$, $J_{e\mu23} \equiv \Im[U_{e2}U_{\mu3}U^{*}_{e3}U^{*}_{\mu2}]$ and $J_{e\mu13} \equiv \Im[U_{e1}U_{\mu3}U^{*}_{e3}U^{*}_{\mu1}]$, along with the $3\nu$ Jarlskog invariant.

The deviation from the CP conservation scenario $J_{\alpha\beta ij}=0$ is found in three results. The best-fit values are -0.0137, -0.0128, and -0.0125, for $J_{e\mu12}$, $J_{e\mu13}$, and $J_{e\mu23}$, respectively. These values are consistent with the best-fit value in the standard $3\nu$ scheme ($J=-0.0148$). In addition, the CP conservation hypothesis is excluded with $\Delta \chi^2 = 1.2(0.8)$ with(without) unitarity.

\subsection{Conditions for $3\nu$ mixing unitarity}\label{sec:result_condition}

\begin{figure}[h!]
\centering 
\includegraphics[width=7cm]{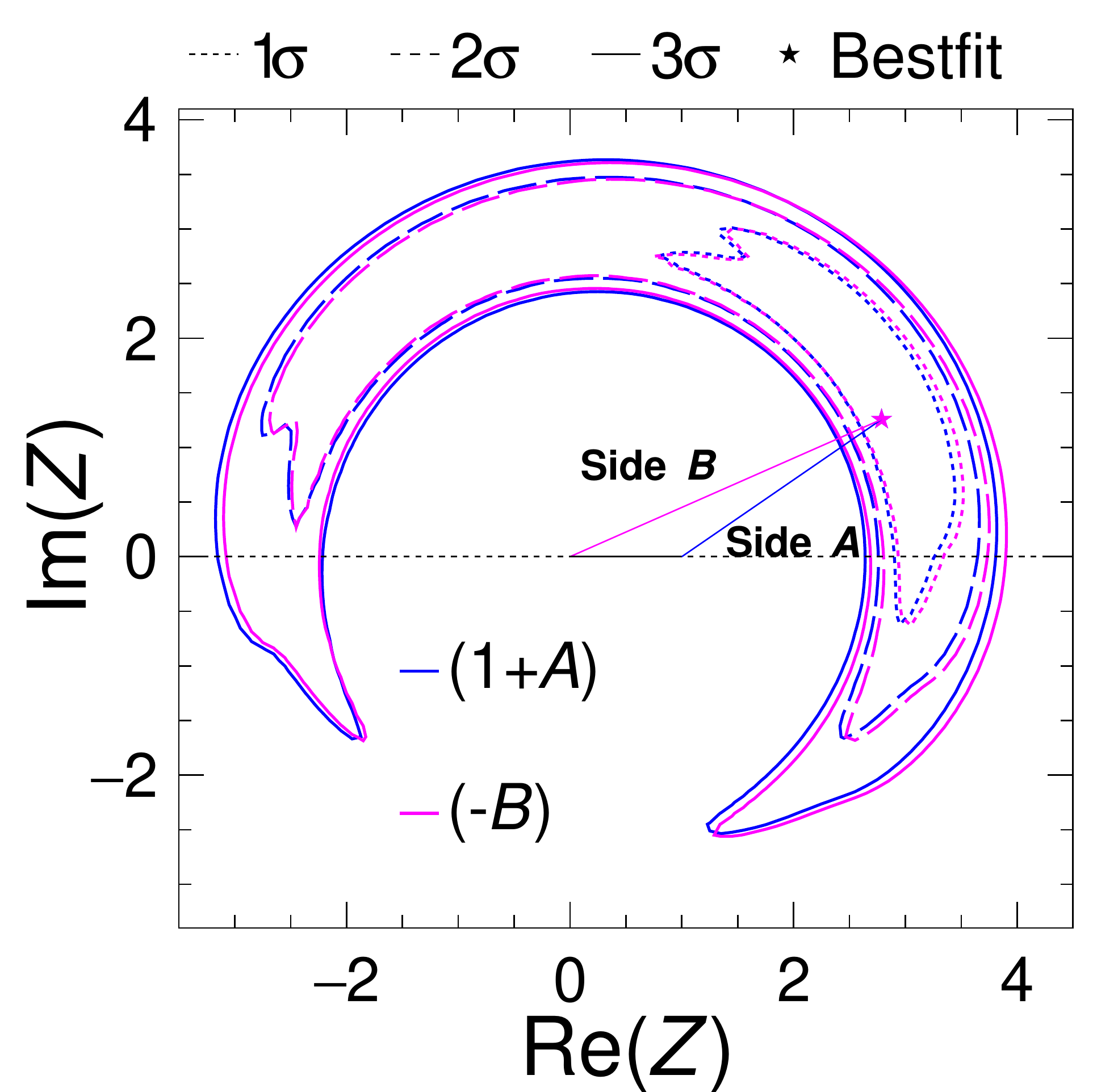}
\includegraphics[width=7cm]{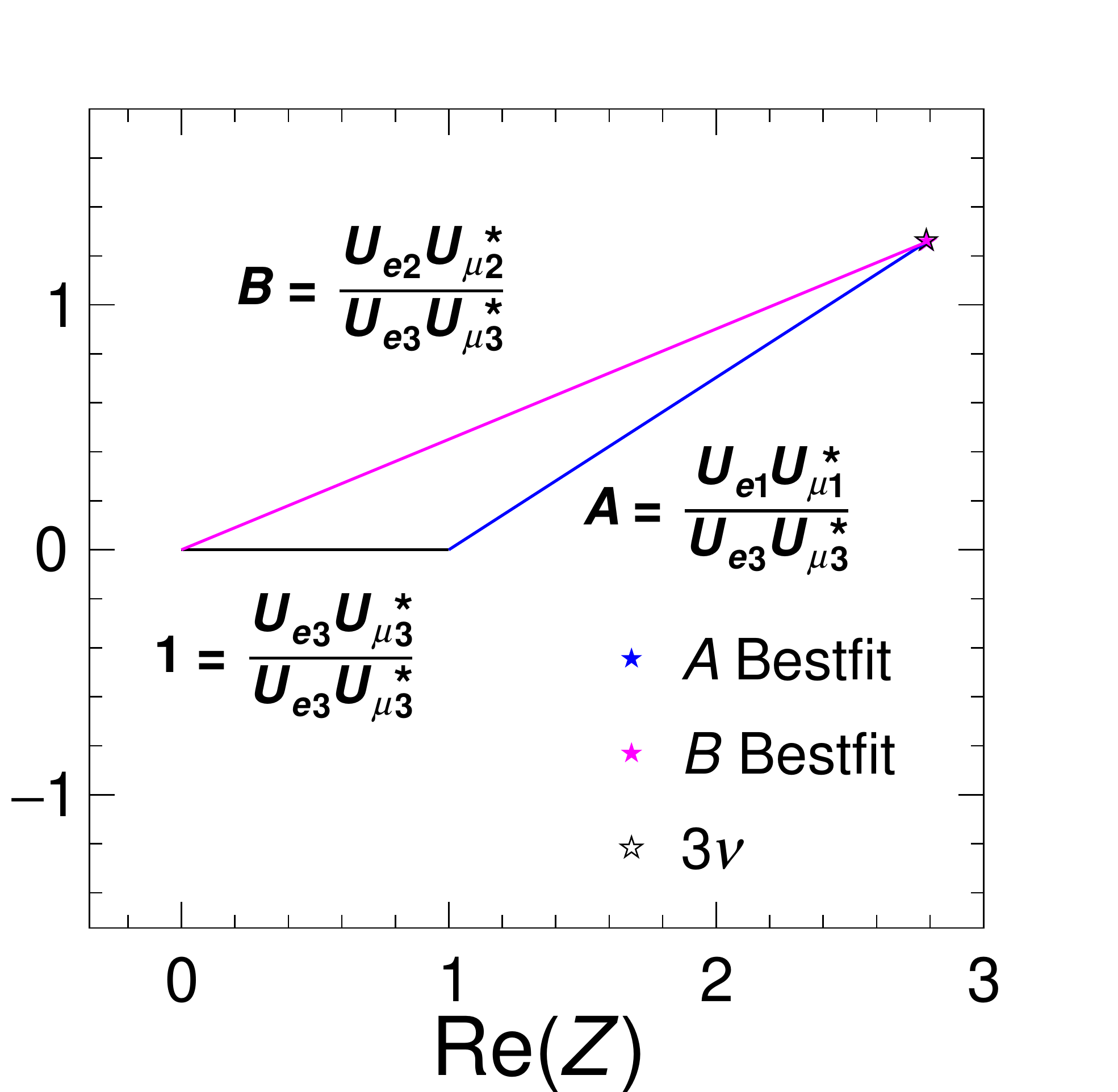}
\caption{$e-\mu$ unitarity triangle without assuming unitarity. Left: allowed regions for the complex numbers $(1+A)=(1+U_{e1}U_{\mu 1}^{*}/U_{e3}U_{\mu 3}^{*})$ (blue) and $(-B)=(-U_{e2}U_{\mu 2}^{*}/U_{e3}U_{\mu 3}^{*})$ (violet). Right: zoom-in of the $e-\mu$ unitarity triangle without assuming unitarity. The triangle consists of three sides $A$, $B$ and $C$, parameterized as in Eq.(\ref{eq:e-mu triangle}). The bestfits of the endpoints of $A$ and $B$ are $(2.7871+1.2573i)$ and $(2.7876+1.2569i)$, shown as blue and violet stars. The bestfit of the vertex of the unitarity triangle assuming unitarity ($2.7858+1.2629i$) is also shown (open star) for comparison. For both panel, the x-axes are the real parts and the y-axes are the imaginary parts. }
\label{fig:Triangle_1}
\end{figure}

The closure of each unitarity triangles are direct tests of the $3\nu$ unitarity hypothesis. In Fig.~\ref{fig:Triangle_1}, we show the the $e-\mu$ unitarity triangle on the left panel, and a zoom-in of the comparison between the best-fits with and without unitarity on the right panel. We present here only the $e-\mu$ unitarity triangle for the same reason as the Jarlskog factor analyses Fig.~\ref{fig:Jarlskog}. The $e-\mu$ unitarity triangle corresponds to the unitarity condition,
\begin{equation}
  U_{e1}U_{\mu1}^{*} + U_{e2}U_{\mu2}^{*} + U_{e3}U_{\mu3}^{*} = 0.
\end{equation}
We parameterize the three sides of the $e-\mu$ unitarity triangle as
\begin{equation}\label{eq:e-mu triangle}
  A \equiv \frac{U_{e1}U_{\mu1}^{*}}{U_{e3}U_{\mu3}^{*}} , ~~~~~~B \equiv \frac{U_{e2}U_{\mu2}^{*}}{U_{e3}U_{\mu3}^{*}} , ~~~~~~C \equiv \frac{U_{e3}U_{\mu3}^{*}}{U_{e3}U_{\mu3}^{*}},
\end{equation}
where each side consists of a combination of $U_{ei}U_{\mu i}^{*}$ and is normalized so that side $C$ is unity. If $U$ is unitary, the three sides form a close triangle with $A+B+C\equiv A+B+1=0$. However, this is not always true when the unitarity conditions are not imposed. There could be three different triangles, depending on which side to be normalized to unity. Here side $C$ is chosen for the ease of calculation, because we did not put phases on $U_{e3}$ and $U_{\mu3}$, as shown in Eq.~(\ref{eq:3nu_non_uni_par}). We plot the allowed regions for the complex numbers $(1+A) \equiv (1+U_{e1}U_{\mu1}^{*}/U_{e3}U_{\mu3}^{*})$ and $(-B) \equiv (-U_{e2}U_{\mu 2}^{*}/U_{e3}U_{\mu 3}^{*})$ on the left panel in Fig.~\ref{fig:Triangle_1}. Contours at $1\sigma$ (dotted), $2\sigma$ (dashed) and $3\sigma$ (solid) confidence levels are shown, for $(1+A)$ (blue) and $(-B)$ (violet). Three straight lines representing the three side $A$ (blue), $B$ (violet) and $C$ (black), are plotted in the same figure. A zoom-in of the resultant $e-\mu$ triangle is plotted on the right panel. The bestfit value for $(1+A)$ (blue star) is $(2.7871+1.2573i)$, while for $(-B)$ (violet star) is $(2.7876+1.2569i)$. The $3\nu$ bestfit (open star) is $(2.7858+1.2629i)$. All of them have non-zero imaginary parts, indicating preferences for CP violation. The endpoints of side $A$ and $B$ look coincident, and the $e-\mu$ triangle is nearly close.

In the following we would like to understand the shapes of the contours. Before discussing the features, we would like to remark on the $e-\mu$ unitarity triangle. The vertex of the $e-\mu$ unitarity triangle is given by 
\begin{equation}
  (-B) \equiv (-U_{e2}U_{\mu 2}^{*}/U_{e3}U_{\mu 3}^{*}) = \sin^2\theta_{12} - \frac{\sin\theta_{12} \cos\theta_{12} \cos\theta_{23}}{\sin\theta_{13}\sin\theta_{23}}e^{i\delta_\text{CP}}.
\end{equation}
One finds the vertex goes along a circle, with the center at $(\sin^{2}\theta_{12}, 0)$ in the complex plane, and the radii being $(\sin\theta_{12} \cos\theta_{12} \cos\theta_{23}) / (\sin\theta_{13} \sin\theta_{23})$. The vertex points towards $(-\infty, 0)$ when $\delta_\text{CP} = 0^{\text{o}}$, and rotates around the center counter-clockwise as $\delta_\text{CP}$ increases. Without the $3\nu$ unitarity assumption, the endpoints of side $A$ and $B$ seem to have the same behaviours as in the case of unitarity. The contours, i.e. the allowed regions of the endpoints, are roughly ring shaped. The ambiguity in the radii of both rings are caused by the \textit{octant} degeneracy when measuring $|U_{\mu3}|$, which we have discussed in Sec.~\ref{sec:result_element}. Currently only $\nu_e$ appearance measurements provide knowledge of CP phases, and therefore result in preferred directions of the $e-\mu$ triangle. In our case, the $e-\mu$ triangle prefers the top-right direction, and the bottom-left parts of the rings are unfavored. More data from running experiments like NOvA and T2K, and future experiments like DUNE and T2HK, help shrinking the contours to pin down the CP violation, and test the unitarity. Alternative methods have been discussed, which proposed to determine the combinations of $U_{ei}U_{\mu i}^{*}$ separately \cite{Farzan:2002ct}.

We also present tests of the unitarity by verifying the following quantities:
\begin{equation}
\begin{aligned}
  &\delta_{\alpha}  = 1 - |U_{\alpha 1}|^{2} - |U_{\alpha 2}|^{2} - |U_{\alpha 3}|^{2},~~\mathrm{for}~\alpha = e,\mu,\tau, \\
  &\delta_{i}  = 1 - |U_{e i}|^{2} - |U_{\mu i}|^{2} - |U_{\tau i}|^{2},~~\mathrm{for}~i = 1,2,3,
\end{aligned}
\label{eqn:normalization_definition}
\end{equation}
and
\begin{equation}
\begin{aligned}
  &\zeta_{\alpha\beta}  = U_{\alpha 1}U^{*}_{\beta 1} + U_{\alpha 2}U^{*}_{\beta 2} + U_{\alpha 3}U^{*}_{\beta 3},~~\mathrm{for}~\alpha, \beta = e,\mu,\tau,~~\alpha\neq\beta, \\
  &\zeta_{ij}  = U_{e i}U^{*}_{e j} + U_{\mu i}U^{*}_{\mu j} + U_{\tau i}U^{*}_{\tau j},~~\mathrm{for}~i, j = 1,2,3,~~i\neq j.
\end{aligned}
\label{eqn:closure_definition}
\end{equation}
The normalization deviations from the $3\nu$ unity condition for each row and column are defined in Eq.~(\ref{eqn:normalization_definition}), and the unitarity triangle closure deviations from zero are defined in Eq.~(\ref{eqn:closure_definition}). If the $3\nu$ unitarity holds, these quantities are zero. 

\begin{figure}[h!]
\centering 
\includegraphics[width=7cm]{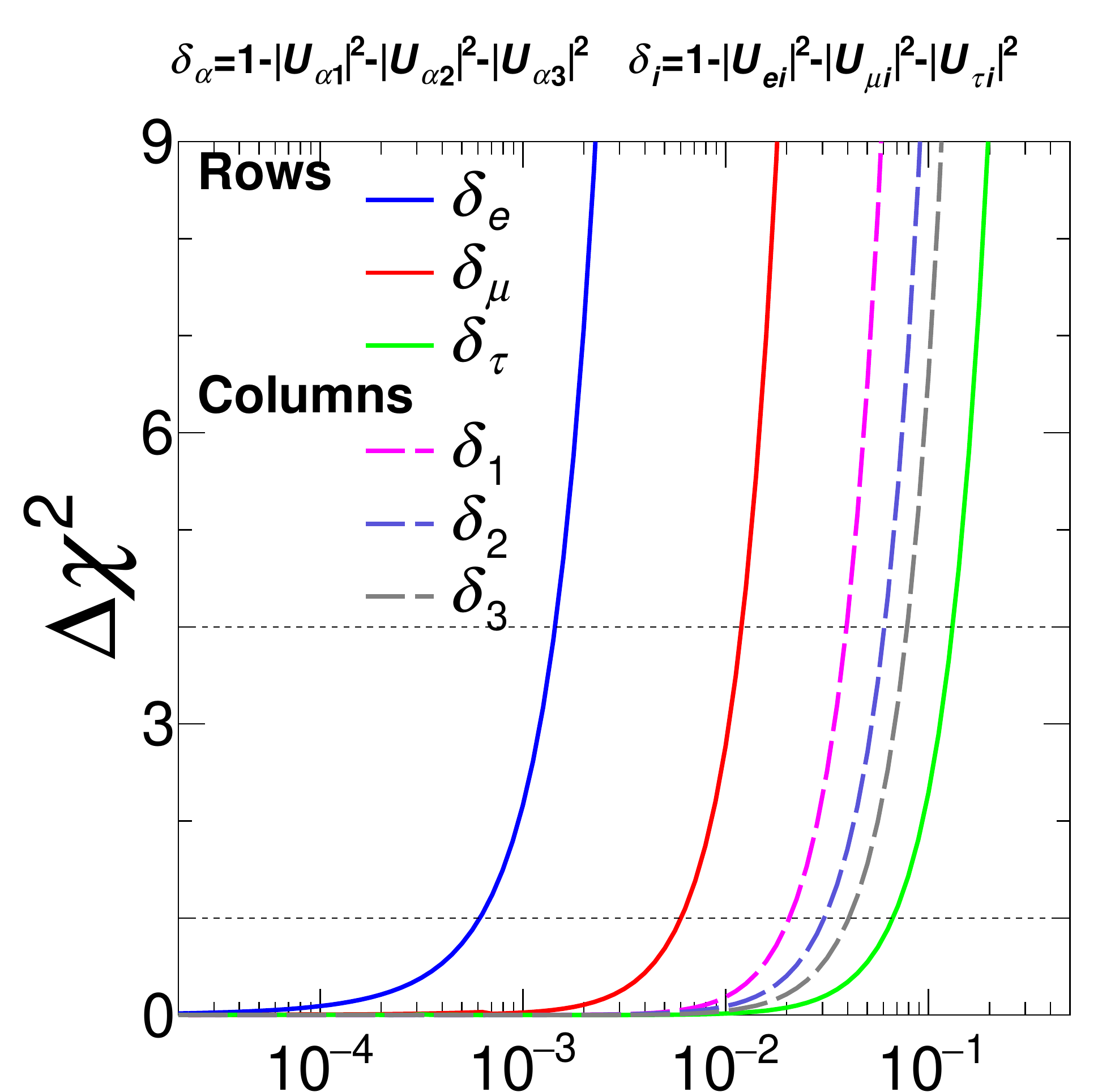}
\includegraphics[width=7cm]{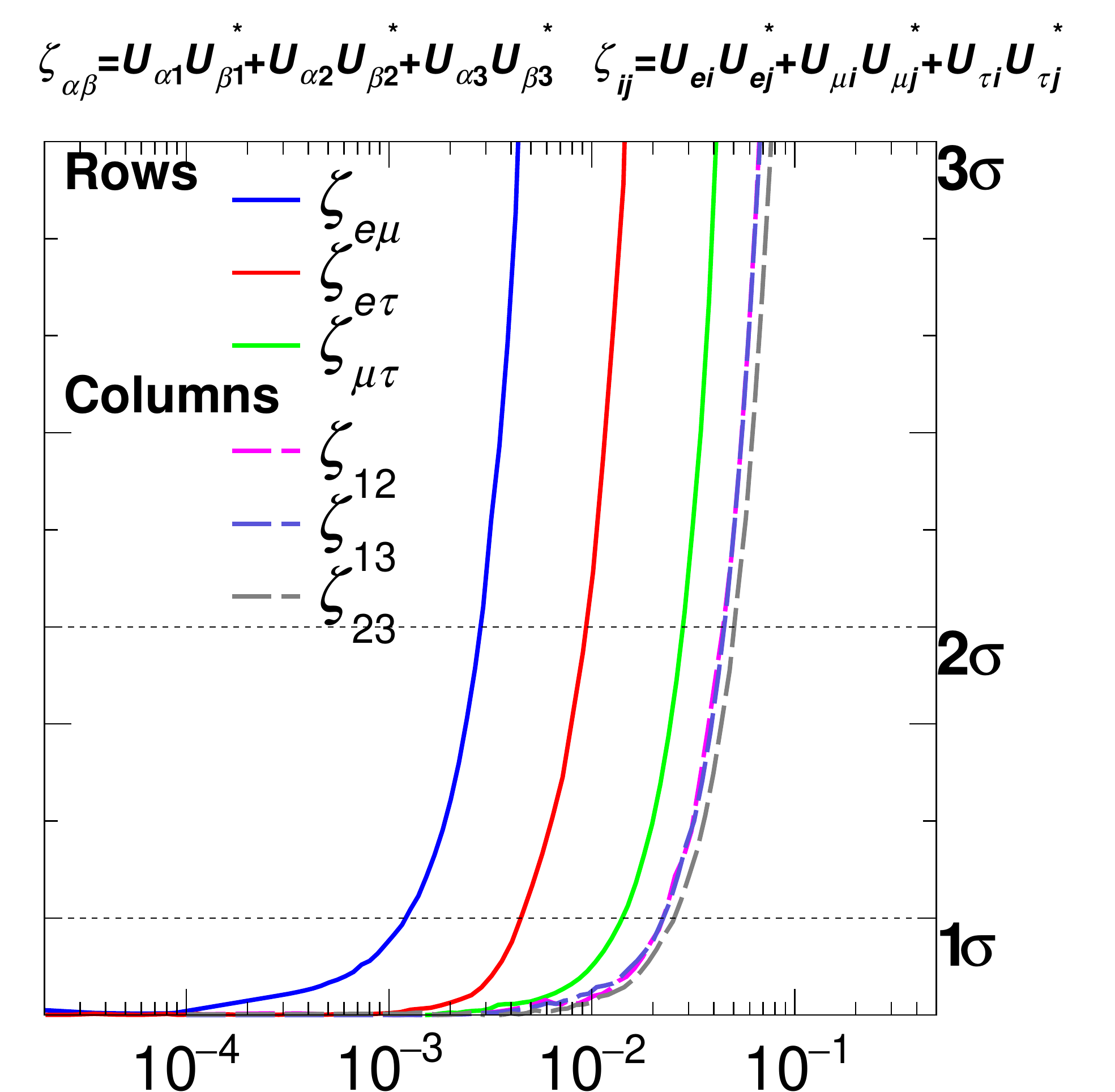}
\caption{Constraints on the non-unitarity. \textbf{Left}: The normalization deviations from unity of each row and column. \textbf{Right}: The deviations of unitarity triangle closure. Any non-zero terms among these quantities indicate non-unitarity. No signal of unitarity violation is found. The curves show upper bounds on the corresponding non-unitary terms. }
\label{fig:unitarity_condition_normalization}
\end{figure}

In Fig.~\ref{fig:unitarity_condition_normalization}, we plot 1-D $\Delta\chi^2$ for $\delta_{\alpha}$($\delta_{i}$) on the left panel, and 1-D $\Delta\chi^2$ for $\zeta_{\alpha\beta}$($\zeta_{ij}$) on the right panel. No signal of non-unitarity is found. We find the violation from unitarity in the $\nu_e$ sector ($\delta_e$) and in the $\nu_{\mu}$ sector $\delta_{\mu}$ well constrained to be less than 0.003 and 0.02, at $3\sigma$ CL. These have benefited from the precise measurements to reactor $\bar{\nu}_e$ disappearance oscillations, accelerator $\nu_{\mu}(\bar{\nu}_{\mu})$ disappearance oscillations, and the solar $^8B$ $\nu_{e}$ flavour conversions. $\delta_{\tau}$ is less known with an upper bound of $\sim0.2$, for we lack of knowledge to the $\nu_{\tau}$ sector. Two out of three elements in a column, $|U_{ei}|$ and $|U_{\mu i}|$, can be measured. Therefore, $\delta_i$ for each column are constrained better than $\delta_{\tau}$, being less than 0.06-0.2 at $3\sigma$ CL. 

The non-closure of the $e-\mu$ unitarity triangle $\zeta_{e\mu}$ is constrained to be less than $\sim0.005$. With little(no) experimental data of the $\nu_{\mu}\rightarrow\nu_{\tau}$($\nu_{e}\rightarrow\nu_{\tau}$) channel, $\zeta_{\mu\tau}$ and $\zeta_{e\tau}$ are known to be $\lesssim0.05$ and $<0.02$, worse than $\zeta_{e\mu}$. The precision measurements in the $\nu_e$ and $\nu_{\mu}$ sectors provide dominant constraints to $\zeta_{e\tau}$ and $\zeta_{\mu\tau}$ via Cauchy-Schwarz inequalities. The remaining three column-wise unitarity triangles, bounded only by the Cauchy-Schwarz inequalities, are known to be $<0.07-0.08$, at $3\sigma$ CL. One can roughly estimate these constraints from $\delta_{1}$, $\delta_{2}$, and $\delta_{3}$, with $\zeta_{ij}\approx\sqrt{\delta_{i}\delta_{j}}$.

\section{Conclusions}\label{sec:conclusion}

The development in neutrino oscillations in the past decades allows us to conduct precision measurements of the neutrino mixing in the active sector ($U^{NU}$). Entering the new precision era, we are able to explore other possibilities, \textit{e.g.} the $3\nu$ unitarity-violated neutrino mixing hypothesis.

An analysis of neutrino oscillations was performed without unitarity assumption in the $3\nu$ picture. We have combined the medium and long-baseline reactor, solar, long-baseline accelerator neutrino data to constrain the mixing matrix in the active sector $U^{NU}$. We have found that elements $U_{ei}$ are measured to be the best among all sectors ($3\sigma$ uncertainty $<20\%$). At the same confidence level, the uncertainties $>20\%$ were obtained in $|U_{\mu i}|$ in the current global analysis. Though currently data for the $\nu_\tau$ sector are limited, via Cauchy-Scharz ineqauilities Eqs.~(\ref{eqn:Cauchy-Schwarz-Sterile_1}) and (\ref{eqn:Cauchy-Schwarz-Sterile_2}) the constraints for this sector can be passed from that in the $\mu$ sector. And, therefore the size of uncertainties for the $\nu_\mu$ and $\nu_\tau$ sectors are similar.
Our result prefers $|U_{\mu 3}|^2>1/2$, which corresponds to the upper-octant solution in the standard $3\nu$ scheme. A negative correlation was noticed between $|U_{\mu1}|$ and $|U_{\mu2}|$, as the $\nu_{\mu}$ disappearance measurements determine the combination $|U_{\mu1}|^2+|U_{\mu2}|^2$. The $\nu_{e}$ appearance measurements help with distinguishing $|U_{\mu1}|$ and $|U_{\mu2}|$ at $1\sigma$ C.L., as well as the "octant-like degeneracy".

Other properties of $U^{NU}$ were further discussed. The CP-violation in the $e-\mu$ sector was investigated by measuring three Jarlskog factors $J_{e\mu12}$, $J_{e\mu13}$, and $J_{e\mu23}$. The result is similar to that in the case of $3\nu$ unitarity, though the uncertainties are slightly worse.
Furthermore, the $e-\mu$ unitary triangle coincides with the unitary $3\nu$ mixing scheme~Fig.~\ref{fig:Triangle_1}. 
This encouraged us to test the assumption of $3\nu$ unitarity Eq.~(\ref{eqn:normalization_definition}) and (\ref{eqn:closure_definition}), as given in Fig.~\ref{fig:unitarity_condition_normalization}. \textbf{Normalizations:} $\delta_e$ and $\delta_\mu$ (Eq.~(\ref{eqn:normalization_definition})) are relatively well-constrained. \textbf{Closure:} we know $\zeta_{e\mu}$ (Eq.~(\ref{eqn:closure_definition})) the best. No significant violations to these conditions are observed. 

Our results also bring some prospects of the future experiments on the $3\nu$ unitarity hypothesis testing, while we summarise the current status in Table~\ref{tab:Experiments_Probability}. The precision of $|U_{e1}|$ and $|U_{e2}|$ can be further improved by the $\bar{\nu}_e$ detection in the JUNO experiment. Future accelerator experiments DUNE and T2HK can provide more data relevant to the $\nu_{\mu}$ sector, and reduce the statistical uncertainties. Further, the large matter density and long baseline of DUNE can be used to measure the NC matter effects due to the active-sterile mixing. The lack of $\nu_\tau$ events brings the large uncertainties of $|U_{\tau i}|$, which might be improved by the possibility of $\nu_\tau$ detection in DUNE.
On the phenomenological analysis itself, we need more information from experiments that is not based on the $3\nu$ assumption, \textit{such as} systematics uncertainties.

We close up this work with a conclusion that we have not found any significant evidence for the $3\nu$ non-unitarity, though the uncertainties (mainly from statistical uncertainties) still needs to be improved. We see the prospect of future DUNE, T2HK, and JUNO data, and are looking forward to any experimental proposals for $\nu_\tau$ oscillation channels. At the mean time, we will keep sharpening our simulation and analysis tools.

\acknowledgments
Jiajie Ling acknowledges the support from National Key R\&D program of China under grant NO. 2018YFA0404013 and National Natural Science Foundation of China under Grant NO. 11775315.
This work was supported in part by Guangdong Basic and Applied Basic Research Foundation under Grant No. 2019A1515012216 and National Natural Science Foundation of China under Grant Nos. 11505301 and 11881240247. Jian Tang acknowledges the support from the CAS Center for Excellece in Particle Physics (CCEPP).
TseChun acknowledges supports from Postdoctoral recruitment program in Guangdong province and the Physics Division, National Center of Theoretical Science of Taiwan. 

\bibliographystyle{JHEP}
\bibliography{paper.bib}

\end{document}